\documentclass[a4paper, 12pt]{report}

\usepackage[utf8]{inputenc}
\usepackage[T1]{fontenc}
\usepackage[english]{babel}

\usepackage{geometry}
\usepackage{setspace}

\usepackage{fancyhdr}
\usepackage{amsmath, amsfonts, amssymb, mathtools, physics, dcolumn}

\usepackage{theorem}

\usepackage[version=4]{mhchem}

\usepackage{array}
\usepackage{tabularx}
\usepackage{multirow}

\usepackage{graphicx}
\usepackage{subcaption}
\usepackage[justification=raggedright]{caption}

\usepackage{hyperref}
\hypersetup{
    colorlinks=true,
    linkcolor=blue,
    citecolor=blue,
    urlcolor=blue,
}

\usepackage{natbib}
\usepackage{xcolor, soul}
\definecolor{mouhablue}{RGB}{87, 74, 226}
\definecolor{myblue}{RGB}{21, 97, 109}
\definecolor{mygreen}{RGB}{0, 100, 0}

\newcommand{\vtheta}{\boldsymbol{\theta}}

\usepackage{makecell}

\usepackage[most]{tcolorbox}
\newcounter{mybox}
\newtcolorbox[use counter=mybox]{bluebox}[2][]{
  title=BOX~\themybox:~#2,
  colback=myblue!10,
  colframe=myblue,
  boxrule=1pt,
  arc=4pt,
  left=6pt, right=6pt, top=6pt, bottom=6pt,
  fonttitle=\bfseries,
  #1
}

\newcounter{algorithms}
\newtcolorbox[use counter=algorithms]{algobox}[2][]{
  title=Algorithm~\thealgorithms:~#2,
  colback=mygreen!10!white,
  colframe=mygreen,
  boxrule=1pt,
  arc=4pt,
  left=6pt, right=6pt, top=6pt, bottom=6pt,
  fonttitle=\bfseries,
  #1
}

\usepackage{algorithm}
\usepackage{algorithmic}



\usepackage{lipsum}
\usepackage{appendix}

\usepackage{booktabs}

\usepackage{chemfig}             

\setchemfig{
  bond length = 9mm,     
  angle increment = 30,  
  atom sep = 0pt
}

\newcommand\arcbetweennodes[3]{%
\pgfmathanglebetweenpoints{\pgfpointanchor{#1}{center}}{\pgfpointanchor{#2}{center}}%
\let#3\pgfmathresult}
\newcommand\arclabel[6][black,-stealth,shorten <=1pt,shorten >=1pt]{%
\chemmove{%
\arcbetweennodes{#4}{#3}\anglestart \arcbetweennodes{#4}{#5}\angleend \draw[#1]([shift=(\anglestart:#2)]#4)arc(\anglestart:\angleend:#2); \pgfmathparse{(\anglestart+\angleend)/2}\let\anglestart\pgfmathresult \node[shift=(\anglestart:#2+1pt)#4,anchor=\anglestart+180,inner sep=0pt,
outer sep=0pt]at(#4){#6};}}
\newcommand\namebond[5][-1pt]{\chemmove{\path(#2)--(#3)node[midway,#4,yshift=#1,black]{#5};}}

\let\oldAA\AA
\renewcommand{\AA}{\text{\normalfont\oldAA}}
\usepackage{siunitx}  
\newcommand{\starlbl}{\textsuperscript{$\ast$}}

\usepackage{slashbox}

\begin{document}




\thispagestyle{empty}
\begin{minipage}{3cm}
\begin{flushleft}
\includegraphics[scale=0.7]{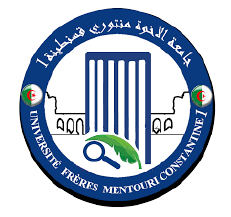}
\end{flushleft}
\end{minipage}
\hfill
\begin{minipage}{9cm}
\begin{center}
\setstretch{1}
\scriptsize{PEOPLE’S DEMOCRATIC REPUBLIC OF ALGERIA\\
MINISTRY OF HIGHER EDUCATION AND SCIENTIFIC RESEARCH\\
UNIVERSITY FRÈRES MENTOURI OF CONSTANTINE 1\\
FACULTY OF FUNDAMENTAL SCIENCES\\
DEPARTMENT OF PHYSICS}\\
\end{center}
\end{minipage}
\hfill
\begin{minipage}{3cm}
\begin{flushright}
\includegraphics[scale=0.7]{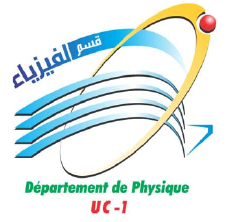}
\end{flushright}
\end{minipage}

\bigskip
\bigskip
\begin{center}
\LARGE \textbf{MASTER DISSERTATION }\normalsize\vspace{\baselineskip}\\
\bigskip

\begin{tabular}{ll}
\small{Domain} & \small{Sciences of Matter}\\
\small{Field} & \small{Physics}\\
\small{Specialty} & \small{Theoretical Physics}\\
&\\
\end{tabular}

\bigskip
\bigskip
\bigskip
\large{THEME :} \normalsize\\
\bigskip
	\rule{0.95 \textwidth}{2pt} \vspace{\baselineskip}\\
		\Large \textbf{FROM VQE TO SQD : MODERN QUANTUM   }
\normalsize\vspace{\baselineskip}\\
		\Large \textbf{ALGORITHMS FOR ELECTRONIC STRUCTURE }
\normalsize\vspace{\baselineskip}\\
		\Large \textbf{PROBLEMS }\normalsize\\
	\rule{0.95 \textwidth}{2pt}	\\ 
\bigskip
\bigskip	
\setstretch{1.2}
Presented by :\\
\textbf{Rabah Abdelmouheymen KHAMADJA}\normalsize\\
\bigskip
Defended on : 06 / 22 / 2025\\
\end{center}
\bigskip
In front of the jury  \\
\begin{center}
\begin{tabular}{lllr}
\textbf{President} & Pr. & K. AIT MOUSSA & University Frères Mentouri. Constantine 1\\
                   &             &                        &\\
\textbf{Supervisor}& Dr. & M.T. ROUABAH & University Frères Mentouri. Constantine 1\\
                   &             &                        &\\
\textbf{Examiner}  & Pr.  & A. BENSLAMA & University Frères Mentouri. Constantine 1\\
                   &             &                        &\\
\end{tabular}
\end{center}


\pagebreak

\chapter*{Dedication}
This work is dedicated:
\begin{itemize}
    \item First to my wonderful parents, my amazing father and my courageous mother, my sister, my extended family, and to my dearest friends L. S. and G. Z. 
    \item Second to the strongest, most faithful, inspiring and truly unbreakable people, the Palestinians. May God have mercy on them.
\end{itemize}

\begin{figure}[H]
    \centering
    \includegraphics[width=0.8\linewidth]{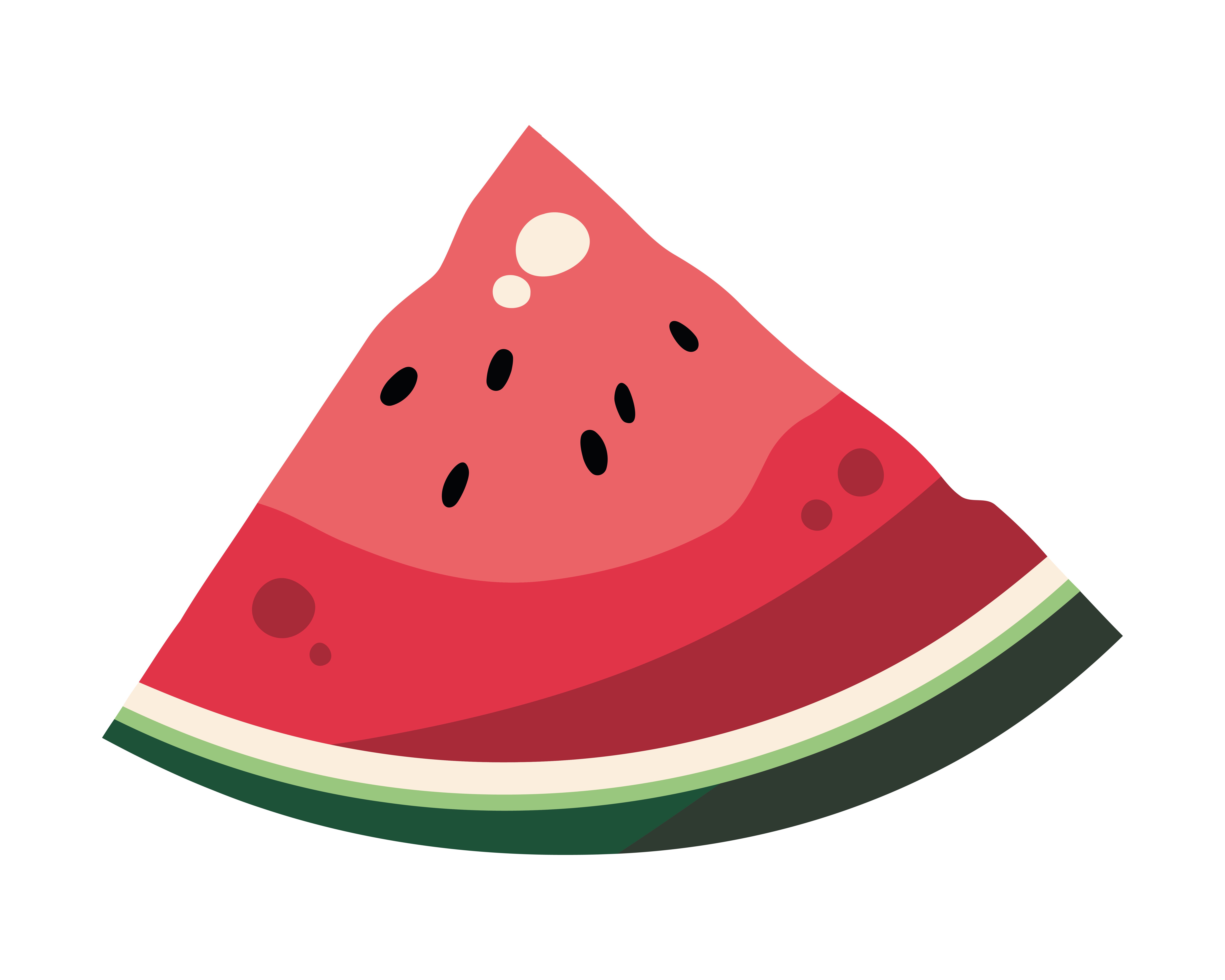}
\end{figure}
\chapter*{Acknowledgments}
First and foremost, I would like to express my deepest gratitude to my mentor, \textbf{Dr. Rouabah}, for the tremendous effort he has invested in me over the past two years. His unwavering belief in my abilities, his constant guidance, and his unparalleled support shaped not only this thesis but also my journey as a researcher. This project simply would not have existed without him.

I would also like to express my gratitude to the entire \emph{Constantine Quantum Technologies} (\textsc{CQTech}) team—\textbf{M.\,T.\,R.}, \textbf{N.\,B.}, \textbf{A.\,T.}, and \textbf{M.\,M.\,L.}—for their mentoring, insightful discussions, and steadfast encouragement throughout this work.

My sincere thanks go to the jury members, \textbf{Prof.\ A.\ Benslama} and \textbf{Prof.\ K.\ Ait Moussa}, for taking the time to review this thesis and for providing constructive feedback that improved the final manuscript.

This document has been produced with the financial assistance of the European Union (Grant no. DCI-PANAF/2020/420-028), through the African Research Initiative for Scientific Excellence (ARISE), pilot programme. ARISE is implemented by the African Academy of Sciences with support from the European Commission and the African Union Commission. The contents of this document are the sole responsibility of the author(s) and can under no circumstances be regarded as reflecting the position of the European Union, the African Academy of Sciences, and the African Union Commission.

Finally, we acknowledge the use of \textbf{IBM Quantum} resources. The experiments presented in this thesis were carried out in part on the IBM Quantum platform; the views expressed here are solely those of the author and do not necessarily reflect the views of IBM or its partners.

\clearpage
\thispagestyle{empty}
\chapter*{Abstract}

Accurately determining electronic ground-state energies is a flagship target for quantum advantage in computational chemistry, yet classical exact methods become intractable as molecular size grows.  
This thesis investigates two recent sampling-based diagonalization algorithms, \textbf{Quantum-Selected Configuration Interaction} (QSCI) and \textbf{Sample-Based Quantum Diagonalization} (SQD), as near-term alternatives to the Variational Quantum Eigensolver (VQE).
Unlike VQE, which suffers from Barren Plateaus and a rapid growth of measurement overhead, QSCI and SQD bypass variational optimization by using a quantum computer only to sample Slater determinants while performing all Hamiltonian diagonalization classically.
A central contribution of this thesis is the first analytical expression for the determinant-discovery step, achieved by establishing a direct analogy between the determinant-discovery step in these algorithms and the classical coupon-collector problem.
This mapping yields an exact expression, as well as a scalable lower-bound estimator, for the expected number of projective measurements required to reveal every determinant contributing to the ground state—thereby quantifying the dominant \emph{sampling bottleneck}.
The analysis is corroborated through a suite of numerical and experimental studies.  
Ideal state-vector simulations, hardware-calibrated noisy simulations, and a real-device run on IBM’s 127-qubit \textsc{IBM Brisbane} processor collectively demonstrate the practical operation of QSCI and SQD across progressively realistic noise environments.  

\section*{Keywords}
Quantum Computing, Quantum Computational Chemistry, Sample-based Quantum Diagonalization
\clearpage

\newgeometry{top=2cm, bottom=2.7cm, left=2.5cm, right=2cm}
\tableofcontents
\restoregeometry
\listoffigures       
\listoftables     

\chapter{Introduction}
\begin{quote}
    
    "\emph{The exact application of these laws leads to equations much too complicated to be soluble}"-- Nobel Prize-winning physicist Paul A.M. Dirac
\end{quote}


\quad One of the fundamental challenges in quantum chemistry involves determining the electronic structure of molecules, particularly finding the ground state solution to the many-body Schrödinger equation. Accurately solving this equation is crucial as it directly impacts numerous practical fields, ranging from drug discovery to advanced materials design, and has significant implications in technology and industry. However, despite its importance, finding exact solutions to the electronic structure problem remains notoriously difficult. As famously stated by Dirac in 1929, the complexity of the equations governing quantum mechanical systems makes them practically unsolvable by conventional analytical methods. Nevertheless, significant efforts have been dedicated to advancing computational chemistry which resulted in a variety of methods capable of approximating solutions. These methods not only include ab initio ones such as Coupled Cluster Theory and M{\o}ller-Plesset perturbative theory, but also semi-empirical ones like Density Functional Theory (DFT) and Density Matrix Renormalization Group (DMRG)~\cite{lewars2010, helgaker2000,Szabo1996, Levine1999}.
Despite remarkable advances, classical electronic structure methods inevitably face a fundamental challenge arising from the exponential growth of the Hilbert space with increasing system size. This complexity originates primarily from the inherently entangled nature of quantum many-body systems, making them exponentially difficult to represent accurately on classical computational resources. Methods such as Full Configuration Interaction (FCI), though highly accurate, quickly become intractable as the size of molecules grows, severely restricting the applicability of exact classical solutions to very small molecular systems. For instance, the biggest implementation of FCI to date involved only $26$ electrons in $23$ orbitals~\cite{gao2024distributed}.  

Quantum computing has recently emerged as a promising pathway to overcome the limitations inherent to classical methods. Since Feynman predicted that we need quantum systems to simulate quantum systems, we have seen the emergence of multiple quantum algorithms that could, theoretically, outperform classical ones~\cite{lipton2021introduction}. The last decade has witnessed significant breakthroughs in quantum hardware~\cite{gyongyosi2019survey}, notably with systems now surpassing the scale of 100 qubits, marking a crucial milestone toward practical quantum advantage. Although current quantum devices, characterized as Noisy Intermediate-Scale Quantum (NISQ)~\cite{preskill2018quantum} computers, still face considerable challenges related to noise and decoherence, they hold significant promise for addressing problems that remain out of reach for even the most powerful classical computers. This rapid progression in hardware, along with the inability to implement the aforementioned algorithms, has motivated the exploration and development of hybrid quantum-classical algorithms specifically tailored to leverage near term quantum resources.

To address the inherent limitations of NISQ devices, Variational Quantum Algorithms (VQAs) have emerged as a practical solution, combining the strengths of classical computation with quantum resources. In this hybrid quantum-classical approach, computational tasks are distributed strategically, leveraging quantum processors for tasks where quantum advantage could manifest, while classical computers handle optimization, post-processing, and error mitigation. A prominent example is the Variational Quantum Eigensolver (VQE)~\cite{peruzzo2014vqe}, an algorithm specifically designed for approximating ground states of Hamiltonians by utilizing trial states prepared via parametrized quantum circuits that are optimized classically. Beyond VQE, several other hybrid algorithms have been proposed, addressing diverse computational tasks such as quantum machine learning, combinatorial optimization, and quantum dynamics~\cite{cerezo2021variational}.\\
Despite the momentum they gained, recent studies have revealed that VQAs suffer a fundamental problem of Barren Plateaus (BPs)~\cite{larocca2025barren}, where the cost function landscape becomes exponentially flat rendering the classical optimization of parameters impossible. Furthermore, the VQE specifically faces an additional roadblock, the measurement problem~\cite{gonthier2022measurements}. The cost of estimating the cost function in VQE, a necessary step, grows rapidly with the system size posing a significant challenge when scaling VQE for sizes beyond a dozen qubits. 

On the other hand, a new class of quantum algorithms for the electronic structure problem had been recently introduced, inspired from the classical Selected Configuration Interaction (SCI), Quantum-SCI (QSCI)~\cite{kanno2023quantum} was proposed as an alternative approach to VQE. Similarly to SCI, in QSCI the quantum computer is used to sample ``configurations`` (Slater determinants) from a prepared approximate ground state that are used to generate a subspace in which the Hamiltonian is projected and diagonalized to approximate the ground state, this approach overcomes the measurement problem facing VQE and promises better scalability. The approximate ground state can be prepared through a VQE routine or multiple other ways such as quantum time evolution~\cite{sugisaki2024hamiltonian,mikkelsen2412quantum}.

Given the current limitations of quantum hardware, a new paradigm appeared, Quantum-Centric Supercomputing (QCSC)~\cite{alexeev2024quantum}, a promising framework aimed to maximize the potential of available quantum resources by closely integrating them with high performance classical computers. The fundamental motivation behind QCSC is to strategically distribute computational tasks, ensuring quantum processors focus exclusively on tasks that best exploit their capabilities, while classical supercomputers handle intensive classical computations, post-processing, and error mitigation tasks. Such an integrated strategy is particularly relevant in fields like materials science, where complex quantum phenomena need accurate modeling beyond classical capabilities~\cite{alexeev2024quantum}. By adopting this approach, QCSC enhances computational scalability and effectively extracts maximum performance from current quantum hardware, paving the way toward practical quantum advantage in the near term. 
A concrete example of this quantum centric framework is the Sample-Based Quantum Diagonalization (SQD) algorithm introduced by Robledo-Moreno \emph{et al.}~\cite{robledo2024chemistry}. This algorithm is built on QSCI, the quantum computer is used to prepare an approximate ground state which is sampled to reveal configurations. These configurations then undergo further classical postprocessing and batching before being used to generate multiple subspaces. SQD workflow involves also an additional error correction scheme~\cite{robledo2024chemistry} that corrects the erroneous configurations based on the ground state resulting from the batch diagonalizations. This method had been tested on the \ce{N2} molecule and iron-sulfur clusters with up to $77$ qubits~\cite{robledo2024chemistry}, marking the largest electronic structure problem solved with a quantum computer to date, it has also been applied to probe supramolecular interactions~\cite{kaliakin2024accurate} and open shell systems~\cite{liepuoniute2024quantum}.

However, the sample based algorithms also face a fundamental challenge concerning their scalability that was initially raised by Reinholdt \emph{et al.}~\cite{reinholdt2025exposing} in their numerical analysis. When sampling from a probability distribution dominated by some determinants, a large number of measurements is required to uncover rare -yet chemically relevant- determinants. This cost grows rapidly with the system size, especially in weakly correlated molecules, putting the scalability of these algorithms in question.

This dissertation contributes to the current literature with:
\emph{(i)~Sampling complexity theory.}  
By proving that the determinant discovery step in Quantum Selected CI and Sample-Based Quantum Diagonalization is formally equivalent to the classical \emph{coupon–collector} problem, providing an exact formula and a scalable lower bound estimator for the required shot count.  
\emph{(ii)~A real implementation of a QSCI/SQD workflow.}  
An experiment with the water molecule, built on \texttt{Qiskit}~2.0 and featuring symmetry filtering, is validated in three regimes: ideal simulation, hardware-calibrated noisy simulation, and execution on IBM’s 127-qubit \textsc{Brisbane} quantum computer, achieving chemical accuracy with orders-of-magnitude fewer shots than VQE.

\subsubsection{Thesis Roadmap}
Chapter 2 surveys the background of NISQ hardware, the electronic-structure problem, and the evolution from VQE to sampling-based methods.  
Chapter 3 develops the coupon–collector framework and analyzes the barren-plateaus and measurement-overhead issues.  
Chapter 4 details the computational setup and compares VQE, QSCI, and SQD across ideal, noisy and hardware regimes.  
Chapter 5 synthesizes the findings, discusses their implications for quantum-centric supercomputing, and outlines future directions.

\chapter{Methodology}
\begin{quote}
    "\emph{Life is nothing but an electron looking for a place to rest.}"-- Nobel Prize-winning physiologist Albert Szent-Györgyi 
\end{quote}
\label{chap2}
\section{Current State of Quantum Computing}
Quantum computers were originally conceived as simulators for other
quantum mechanical systems, an idea traceable to Feynman’s observation that
classical resources grow exponentially when modeling quantum dynamics.  For
electronic structure theory, this wall manifests in the
\(\mathcal{O}(\binom{2N}{n_{el}})\)\footnote{$N$ is the number of spatial orbitals included in the active space, $n_{el}$ is the number of electrons in the active space.} scaling of
Full Configuration Interaction (FCI) and similarly prohibitive costs in high‐order
coupled‐cluster or Selected Configuration Interaction (SCI) expansions.  Quantum algorithms promise a
drastic reduction: Lloyd showed that generic local Hamiltonian dynamics can be
simulated with polynomial resources~\cite{lloyd1996universal}, and Abrams \& Lloyd~\cite{abrams1997simulation}, followed by
Aspuru‐Guzik \emph{et al.}~\cite{aspuru2005simulated}, translated this result into an explicit scheme for
computing molecular energies by mapping second‐quantized fermionic operators to
qubits and applying quantum phase estimation (QPE).  Subsequent refinements
lowered the gate complexity from \(\mathcal{O}(N^4)\) to
\(\mathcal{O}(N^3)\) as in low‐rank factorizations~\cite{motta2021low} and qubitization~\cite{low2019hamiltonian}, while
demonstrations of nitrogen fixation~\cite{reiher2017elucidating} and excitation spectra indicate that a
fully error‐corrected device with a few thousand logical qubits could rival the
best classical methods.  The practical realization of this vision, however,
depends on hardware that is not yet available, leading to the present focus on
the \emph{Noisy Intermediate‐Scale Quantum}~\cite{preskill2018quantum} regime described below.

\subsection{The NISQ Era}
The present generation of quantum processors defines the NISQ era: devices comprising only a few-dozen to
a few-thousand physical qubits that suffer from non-negligible control errors
and finite coherence times.  Typical single- and two-qubit gate infidelities
lie in the range \(10^{-3}\)–\(10^{-2}\), while relaxation and dephasing times
\(T_{1},T_{2}\) rarely exceed a few tens of microseconds.  These figures place
a practical ceiling on the circuit depth \(d_{\text{max}}\) that can be
executed before decoherence erases quantum advantage.  Full
fault-tolerant error correction remains out of reach because it would demand
several orders of magnitude more physical qubits per logical qubit than are
currently available.  Consequently, algorithm design in the NISQ regime
emphasizes error-\emph{mitigation} rather than complete error correction and favors hybrid quantum–classical workflows such as variational quantum
algorithms (VQAs), notably the variational quantum eigensolver (VQE), the
quantum {approximate} optimization algorithm (QAOA), and their adaptive
variants. Ans\"atze circuits should respect the hardware’s coupling graph to avoid excessive \textsc{swap} overhead. With recent advances in quantum processing units, a \emph{Quantum-Centric Supercomputing} (QCSC)~\cite{alexeev2024quantum} paradigm has emerged, in which a NISQ-scale processor is tightly integrated with a classical high-performance computing backend; early demonstrations indicate that this hybrid architecture can extend the reach of quantum chemistry and optimization workloads while remaining within today’s hardware limits \cite{robledo2024chemistry}.

\section{Quantum Computational Chemistry}

Predicting molecular energies and properties with chemical accuracy is crucial in various industrial fields such as catalysis, drug discovery, and materials design.  While
classical \emph{ab initio} techniques have delivered remarkable successes,
their computational cost grows steeply with both system size and correlation
strength, placing many scientifically and industrially relevant targets out
of reach.  Quantum computers, by natively processing superposition and
entanglement, offer a route to bypass this exponential wall.
The material that follows presents the electronic structure problem, tracing the derivation of the electronic Hamiltonian, its
second-quantized form, and the mappings required to run quantum algorithms on
present-day hardware.

\subsection{The Electronic Structure Problem}
In atomic units the full molecular Hamiltonian is
\begin{align}
  \hat H_{\mathrm{mol}}
    = -\sum_{m} \frac{\nabla_m^{2}}{2M_m}
      -\sum_{i} \frac{\nabla_i^{2}}{2}
      -\sum_{m,i} \frac{Z_m}{|\mathbf{R}_m-\mathbf{r}_i|}
      +\sum_{m<n} \frac{Z_mZ_n}{|\mathbf{R}_m-\mathbf{R}_n|}
      +\sum_{i<j} \frac{1}{|\mathbf{r}_i-\mathbf{r}_j|}.
  \label{eq:molHam}
\end{align}
where $\mathbf{R}_m$ and $M_m$ ($\mathbf{r}_i$ and $m_i$) are, respectively, the position vector and the mass of the $m$'th nucleus ($i$'th electron) in the system, $Z_m$ is the atomic number (describing the charge) of the $m$'th nucleus.\\
\textbf{Born Oppenheimer Approximation} \cite{Born1927,Szabo1996}: since the nuclei are much heavier than electrons, we simplify the problem by considering them fixed in a given position (\(\nabla_m^{2}/(2M_m)\to0\)), the nucleus-nucleus term becomes a constant \(E_{\mathrm{nuc}}\).  
The electronic Hamiltonian is then given by:
\begin{equation}
  \hat H_{\mathrm{el}}
    = -\sum_{i} \frac{\nabla_i^{2}}{2}
      -\sum_{m,i} \frac{Z_m}{|\mathbf{R}_m-\mathbf{r}_i|}
      +\sum_{i<j} \frac{1}{|\mathbf{r}_i-\mathbf{r}_j|}+ E_{\mathrm{nuc}}.
  \label{eq:firstElHam}
\end{equation}

The central task of quantum chemistry is to obtain the ground-state energy
\(E_0=\min_{\Psi}\langle\Psi|\hat H_{\mathrm{el}}|\Psi\rangle\) and related
properties (forces, dipole moments, excitation gaps) to \emph{chemical
accuracy}, conventionally defined as \(\leq\!1\;\mathrm{kcal\,mol^{-1}}
\approx 1.6\;\mathrm{mHa}\)(milliHartree).
Achieving this tolerance is essential for quantitative predictions of
bond-breaking energies, reaction barriers, and spectroscopic line
shifts, yet classical exact diagonalization becomes prohibitive once the
number of spin-orbitals exceeds \(\sim\!20\).


In quantum chemistry calculations, the Hamiltonian is projected into a finite set of basis wave functions describing molecular orbitals. \emph{Atomic orbital basis sets} are used to expand molecular orbitals as linear combinations of atomic-like functions. A basis set is a chosen set of functions (basis functions) that approximate the atomic orbitals for the atoms in the molecule. \emph{Minimal basis sets} are the simplest choice: in a minimal basis, there is only one basis function for each atomic orbital (typically each core and valence orbital) of each atom. For example, a minimal basis for a first-row atom includes just the 1s orbital, whereas for a second-row atom it includes the 1s orbital (first shell) and the 2s and 2p orbitals (second shell) as basis functions.  

These basis functions are often taken to be \emph{Gaussian-type orbitals (GTOs)}---functions with Gaussian radial dependence---because Gaussians allow efficient analytical evaluation of the many required integrals. In contrast, the true atomic orbitals (hydrogen-like solutions) are \emph{Slater-type orbitals (STOs)} with an exponential decay $\sim e^{-\zeta r}$; while STOs are more physical, they are less convenient computationally. The popular compromise is to approximate STOs by linear combinations of a few Gaussian functions. For instance, the \emph{STO-nG} family of basis sets (Slater-type orbital – $n$ Gaussian) uses $n$ primitive Gaussians per orbital to mimic a single Slater orbital. A common minimal basis choice is \emph{STO-3G}, which uses 3 Gaussians for each atomic orbital~\cite{helgaker2000,Szabo1996}.

\subsection{Second Quantization and Mapping to Qubits}

Having defined a set of spin orbitals, we switch to the second-quantization (particle-hole) formalism. In this formalism, each spin orbital corresponds to a fermionic mode that can be either unoccupied or occupied by an electron . We introduce \emph{creation} and \emph{annihilation} operators, $\hat{a}_p^\dagger$ and $\hat{a}_p$, which act on these modes. These obey the \emph{canonical anticommutation relations}:
\[
\{a_p,\;a_q\} = 0, \qquad \{a_p^\dagger,\;a_q^\dagger\} = 0, \qquad \{a_p,\;a_q^\dagger\} = \delta_{pq}.
\]
These relations ensure the correct fermionic statistics: for example, $(a_p^\dagger)^2 = 0$ enforces the Pauli exclusion principle.

In fermionic creation and annihilation operators, the electronic Hamiltonian
takes the form
\begin{equation}
  \hat H_{\mathrm{el}}
    = \sum_{p,q} h_{pq}\,a_p^{\dagger} a_q
      + \tfrac12 \sum_{p,q,r,s} h_{pqrs}\,a_p^{\dagger} a_q^{\dagger} a_s a_r,
  \label{eq:secondElHam}
\end{equation}
with integrals
\begin{align}
  h_{pq} &=
    \int \psi_p^{*}(\mathbf{r})
          \Bigl(-\tfrac12\nabla^{2}-\sum_m\frac{Z_m}{|\mathbf{R}_m-\mathbf{r}|}\Bigr)
          \psi_q(\mathbf{r})\,d\mathbf{r}, \label{eq:hpq} \\
  h_{pqrs} &=
    \int \frac{\psi_{p}^{*}(\mathbf{r}_1)\psi_{q}^{*}(\mathbf{r}_2)
               \psi_{r}(\mathbf{r}_1)\psi_{s}(\mathbf{r}_2)}
              {|\mathbf{r}_1-\mathbf{r}_2|}\,
          d\mathbf{r}_1 d\mathbf{r}_2. \label{eq:hpqrs}
\end{align}

To simulate fermions on a quantum computer, we must map fermionic operators to qubit operators. The \emph{Jordan–Wigner (JW) transformation}~\cite{JordanWigner1928} provides such a mapping while preserving the anticommutation relations. For $N$ modes indexed by $p = 0, \dots, N-1$, we define:
\[
a_p \mapsto \frac{1}{2}(X_p + iY_p) \bigotimes_{j=0}^{p-1} Z_j, \qquad
a_p^\dagger \mapsto \frac{1}{2}(X_p - iY_p) \bigotimes_{j=0}^{p-1} Z_j,
\]
where $X_p, Y_p, Z_p$ are the Pauli operators on qubit $p$ and the product of $Z$ operators is known as the Jordan–Wigner string.

The JW mapping encodes orbital occupations directly and
locally but generates Pauli strings whose weight scales linearly \(\mathcal{O}(N)\) with the
number of spin‐orbitals leading to deep circuits. The \emph{parity}
mapping likewise has linear weight yet stores parity information explicitly,
which simplifies certain symmetry‐tapering procedures~\cite{seeley2012bravyi}. A
more locality‐friendly alternative is the \emph{Bravyi–Kitaev} transformation,
whose maximal Pauli weight grows only as \(\mathcal{O}(\log_2 (N))\) by mixing
occupation and parity data; this usually shortens circuit depth and lowers
two‐qubit gate counts in Trotter steps or qubitization kernels
\cite{seeley2012bravyi, bravyi2002fermionic,tranter2015b,setia2018bravyi,tranter2018comparison}, although it does not automatically yield greater noise resilience \cite{sawaya2016error}.  Even lower asymptotic weight,
\(\mathcal{O}\!\bigl(\log_{3}(2N)\bigr)\), can be achieved with the optimal
general encoding on ternary trees proposed in Ref.~\cite{jiang2020optimal}.
After any such mapping the electronic Hamiltonian assumes the form of a weighted sum of Pauli strings:

\begin{equation}
  \hat{H}=\sum_{i} w_i P_i,\qquad P_i\in\{\mathbb{I},X,Y,Z\}^{\otimes N}
  \label{eq:Pauli_sum}
\end{equation}

Exploiting conserved quantum numbers (particle number, total spin
\(S^{2}\), \(S_{z}\), and molecular point‐group irreps) permits
qubit tapering~\cite{bravyi2017tapering}, using symmetries to reduce the number of qubits and shrink the operator
list.  Finally, commuting Pauli strings can be partitioned into common
measurement bases, often reducing the term count, a crucial optimization for NISQ‐era experiments
irrespective of the specific quantum‐chemistry algorithm employed.

\subsection{Quantum Algorithms for Quantum Chemistry}
\label{sec:qc-algorithms}

Quantum Phase Estimation (QPE) and Hamiltonian‐simulation techniques offer
provable speed-ups for electronic‐structure problems, but only when run on
fully error-corrected hardware \cite{abrams1999quantum,low2019hamiltonian}.
Current devices cannot supply the millions of high-fidelity logical gates or
the very high number of physical qubits per logical qubit that such
fault-tolerant implementations demand.  Deep Trotter or qubitization
sequences quickly exceed the coherence budget, and logical error rates remain
orders of magnitude above those required for chemically accurate phase
estimation.

These hardware constraints have shifted attention toward \emph{near-term}
NISQ algorithms that operate with
shallow circuits, modest qubit counts, and error-mitigation rather than full
error correction.  The remainder of this chapter introduces NISQ-compatible
approaches tailored to quantum chemistry.  In
Section~\ref{sec:VQAs} Variational Quantum Algorithms are introduced with a focus on the Variational Quantum
Eigensolver (VQE), while in Section~\ref{sec:Qsubspace} sample based subspace search algorithms for quantum chemistry and their recent developments are introduced.



    \section{Variational Quantum Algorithms}
    \label{sec:VQAs}

    \quad Variational Quantum Algorithms (VQAs) are a class of hybrid
quantum–classical algorithms that have received significant attention
within the scientific community as promising candidates to deliver near-term
quantum advantage. These algorithms are considered the quantum
analog of highly successful machine-learning methods; they rely on a classical computer to train a set of parameters ($\vtheta$) used to prepare a trial state $|\psi(\vtheta)\rangle$ that minimizes a cost function measured efficiently on a quantum processor. By leveraging the well-established classical optimizer toolbox~\cite{jones2024benchmarking,tilly2022variational}, and by off-loading the computational burden to the
classical side, VQAs have seen a recent surge of interest. A
description of the common skeleton shared by all VQAs is provided in
Algorithm~\ref{algo:VQAs}. The next sections detail the specific VQAs relevant
to the electronic structure problem, namely the Variational Quantum
Eigensolver (VQE) and its variants, while Box~\ref{box:VQAs} summarizes other
applications of VQAs~\cite{cerezo2021variational}.
\subsection{The Variational Quantum Eigensolver (VQE)}  
Introduced in 2014 by Peruzzo et al.~\cite{peruzzo2014vqe}, the Variational Quantum Eigensolver (VQE) is widely used in quantum chemistry to estimate the ground state energy of electronic structure problems, primarily due to its compatibility with current quantum hardware. VQE is a type of variational quantum algorithm (VQA), as outlined in Algorithm~\ref{algo:VQAs}, where the cost function \( C(\vtheta) \) is defined as the expectation value of the system's Hamiltonian\footnote{This corresponds to Equation~\eqref{eq:cost_fn} with \( \{O_k\} = \{\hat{H}\} \), \( \{\rho_k\} = \{|\phi_0\rangle \langle \phi_0|\} \), and \( \{\lambda_k\} = \{1\} \).}.

  \begin{equation}
  \label{eq:VQE_energy}
      C(\vtheta) = E(\vtheta) =  \langle\psi(\vtheta)|\hat{H}|\psi(\vtheta)\rangle
  \end{equation}
  where $|\psi(\vtheta) \rangle = U(\vtheta) | \phi_0\rangle $ is the state resulting from applying the unitary (ansatz) $ U(\vtheta)$ to an initial state $|\phi_0\rangle$.
Due to its simplicity, VQE serves as a flexible framework that naturally accommodates the choice of various components, making it highly adaptable to different problem settings. In particular, a wide range of classical optimization algorithms can be employed to minimize the cost function~\cite{jones2024benchmarking, tilly2022variational}, from gradient-free methods like Nelder-Mead~\cite{peruzzo2014vqe} to more sophisticated gradient-based or stochastic optimizers, depending on the problem landscape and noise characteristics. 
    
Similarly, the choice of the ansatz $ U(\vtheta) $ is crucial and problem-dependent: chemically inspired ansätze such as the Unitary Coupled Cluster Singles and Doubles (UCCSD) are often used for molecular systems, while hardware-efficient ansätze are preferred for near-term devices due to their shallow circuit depth \cite{belaloui2024ground}. 
Because the cost function is the Rayleigh quotient of \(H\), the variational principle
guarantees \(E(\boldsymbol{\theta}) \ge E_0\), with equality only at the exact
ground state.  In practice one terminates the optimization when successive
energy estimates differ by less than a problem-defined threshold or when the gradient norm falls below a preset tolerance.
For quantum evaluation of the cost function, we make use of the qubit form of the electronic Hamiltonian from Equation \eqref{eq:Pauli_sum},
so that \(E(\boldsymbol{\theta})=\sum_i w_i\langle P_i\rangle\).
Expectation values are estimated from repeated circuit executions (“shots”). In the presence of noise, this process might violate the variational principle.
Although the number of terms scales as \(\mathcal{O}(N^{4})\) with the number of spin-orbitals, commuting‐set grouping
can reduce the effective measurement load; a dedicated discussion to this appears in Chapter \ref{chap3}.
Circuit depth depends on the ansatz: hardware-efficient ansätze's depth scales as
\(O(d_l\times l)\), $d_l$ being the depth of one layer of the ansatz, and $l$ the number of layers (repetitions) of the ansatz, more repetitions allow better accuracy at the cost of higher levels of noise due to deep circuits and a worse trainability. Whereas chemically inspired ansätze such as UCCSD grow with the number of excitation operators considered~\cite{lee2018generalized}.

Initializing the register in the Hartree–Fock determinant provides a
chemically motivated starting point that often accelerates convergence.
Noise can be alleviated through error mitigation techniques such as zero-noise extrapolation, symmetry
verification, and probabilistic error cancellation
\cite{cai2023quantum}.  These techniques fit naturally into
the VQE loop at the cost of making the runtime of the algorithm longer, and in some cases violating the variational principle.
For a comprehensive overview of the various advancements and adaptations of VQE, we refer the reader to recent review articles such as Tilly et al.~\cite{tilly2022variational}.

    \begin{algobox}[label=algo:VQAs]{Structure of a VQA}
    Variational Quantum Algorithms (VQAs) follow a hybrid quantum--classical loop where a parameterized quantum circuit $U(\boldsymbol{\theta})$ is evaluated on quantum hardware to estimate a cost function, and a classical optimizer updates the parameters $\boldsymbol{\theta}$ to minimize this cost, essentially solving for $\vtheta^* = \arg \min_{\vtheta} C(\vtheta)$.
    
    \medskip
    
    \textbf{Cost Function.} Given input states $\{\rho_k\}$ and observables $\{O_k\}$, the cost function is typically defined as
    \begin{equation}
    \label{eq:cost_fn}
    C(\vtheta) = \sum_k \lambda_k \, \text{Tr}\left[ O_k \, U(\vtheta) \rho_k U^\dagger(\vtheta) \right]
    \end{equation}
    where $\{\lambda_k\}$ are task-dependent weights (Lagrange multipliers).
    
    \medskip
    
    \textbf{Ansatz.} The ansatz $U(\boldsymbol{\theta})$ is a parameterized quantum circuit whose structure encodes the optimization landscape. Famous examples include:
    \begin{itemize}
        \item \textbf{Hardware-Efficient Ansatz (HEA):} HEA is a generic name used to call a family of ans{\"a}tze that is tailored to consider the hardware's connectivity and basis gates, first introduced in reference~\cite{kandala2017hardware}. HEAs are mainly a hardware efficient unitary $U(\vtheta_l)$ that is repeated $l$ times, this family of ans{\"a}tze is general and not problem specific, less accurate than problem inspired ones, but for a high number of repetitions $l$ can achieve similar accuracy but might result in a worse trainability (which will be discussed later).
        \item \textbf{Unitary Coupled Cluster (UCC) ansatz:} The UCC ansatz~\cite{lee2018generalized} is the most famous example of problem-inspired ans{\"a}tze, widely used in quantum chemistry electronic structure problems.
        \item \textbf{Quantum Alternating Operator Ansatz:} First introduced as part of the \emph{Quantum Approximate Optimization Algorithm}~\cite{farhi2014qaoa}, holding the same acronym as the algorithm (QAOA), it has an alternating structure sequentially applying a problem unitary and a mixing unitary~\cite{hadfield2019quantum}. This ansatz has been proven to be universal for a specific set of problems, the proof had been generalized for ans{\"a}tze defined by sets of graphs and hypergraphs \cite{morales2020universality}.
        \item \textbf{Hamiltonian Variational Ansatz:} Similar to the QAOA ansatz, this ansatz~\cite{wecker2015progress} is the most intuitive problem-inspired ansatz. It prepares the ground state of a given Hamiltonian by Trotterizing an adiabatic state preparation process where each Trotter step is parametrized and considered a unitary.
    \end{itemize}
    There exists a wide variety of ans{\"a}tze outside the examples mentioned here, each offering trade-offs between expressibility, trainability, and circuit depth.
    \end{algobox}


      \begin{bluebox}[label=box:VQAs]{Applications of VQAs}
      The spiking interest in quantum computing and the pursuit of near term quantum advantage yielded a wide variety of variational algorithms for various applications.\\
      \textbf{Quantum Machine Learning}\\
      With the rising popularity of quantum computing, Quantum Machine Learning (QML) appeared as a promising field. Multiple variational algorithms have been introduced in each stage of a QML algorithm, variational quantum classifiers were introduced as supervised machine learning models \cite{farhi2018classification, schuld2020circuit}, variational quantum autoencoders \cite{romero2017quantum} for efficient compression of data for QML, variational quantum generators \cite{romero1901variational} as a quantum version of Generative Adversarial Networks (GAN), and variational quantum clustering \cite{bermejo2023variational}.\\       
      \textbf{Ground and Excited States}\\
      The ground (and excited) state problem is the most interesting problem for quantum advantage, it is only natural that it has gained its fair share of variational algorithms. The main algorithms for finding ground and excited states using variational quantum approaches include the Variational Quantum Eigensolver (VQE) \cite{peruzzo2014vqe}, Orthogonality Constrained VQE (Variational Quantum Deflation of States) \cite{higgott2019ocvqe}, Subspace Expansion Method \cite{mcclean2017hybrid}, Subspace VQE \cite{nakanishi2019subspace}, Multistate Contracted VQE \cite{parrish2019quantum}, Adiabatically Assisted VQE \cite{garcia2018aavqe}, and Accelerated VQE using $\alpha$-QPE \cite{wang2019accelerated, wang2021minimizing}.\\
      \textbf{Optimization}\\
      Classical optimization problems are usually NP hard due to their combinatorial nature, rendering them an attractive candidate to tackle with quantum computers. The Quantum Approximate Optimization Algorithm (QAOA), since its introduction in 2014 \cite{farhi2014qaoa}, had been applied to a lot of classical optimization problems like max-cut \cite{wang2018qaoamax} and constraint satisfaction problems \cite{lin2016qaoa}, eventually gaining traction in finance with the portfolio optimization problem \cite{brandhofer2022qaoaportfolio}.\\      
      \textbf{Mathematics}\\
      Since most quantum algorithms that treat mathematical problems -such as Shor's factoring algorithm \cite{shor1999polynomial}, and the Harrow–Hassidim–Lloyd (HHL) quantum algorithm \cite{harrow2009quantum} that solves linear systems of equation- require fault tolerant quantum computers, multiple VQAs have been proposed to solve mathematical problems on NISQ devices. They consist of translating the problem's solution to the ground state of a Hamiltonian and variationally preparing it. The Variational Quantum Linear Solver \cite{bravo2023variational, huang2021near, xu2021variational} was introduced as a near-term alternative for the HHL algorithm in solving linear systems of equations, the Variational Quantum Factoring \cite{anschuetz2019variational} algorithm as an alternative to Shor's factoring algorithm, Matrix-Vector multiplication has also been tackled with a variational algorithm \cite{xu2021variational} as well as Non-Linear equations \cite{lubasch2020variational} and Principal Component Analysis (PCA) \cite{larose2019vqsd, cerezo2022vqse}.
      
      \end{bluebox}


\section{Quantum Subspace Methods for Electronic Structure}
\label{sec:Qsubspace}
The electronic‐structure problem scales combinatorially: an exact
full‐configuration–interaction (FCI) treatment of $N_{\mathrm{elec}}$ electrons in $M$ molecular orbitals requires diagonalizing a
Hamiltonian in a space of dimension \(\binom{2M}{N_{\mathrm{elec}}}\), which is
intractable for most chemically relevant molecules. To date, the biggest FCI calculation ever carried involved $26$ electrons in $23$ orbitals~\cite{gao2024distributed}, performed using $256$ servers, including $1.3$ trillion determinants (configurations).  To preserve FCI‐level accuracy at lower cost, \emph{post‐FCI} methods restrict the
calculation to a small subspace within the Hilbert space.  Subspace diagonalization algorithms implement this strategy in various ways, some of which are mentioned in Box~\ref{box:CSM}, once a subspace of dimension $d$ $\{ |\Phi_i\rangle\}_{i=0}^{d}$ is designated, the Hamiltonian is projected onto this subspace to obtain a reduced Hamiltonian $\hat{H}_p$. The ground state and its energy are then obtained by solving the generalized eigenvalue problem with the projected Hamiltonian in the resulting compact basis:
\begin{align}
\label{eq:gen_eig}
\hat{H}_p\,\mathbf{c}&=E_p\,\mathbf{S}\,\mathbf{c},\\
\text{where}\;
(\hat{H}_p)_{ij}=\langle\Phi_i|\hat H|\Phi_j\rangle
\;\text{and}\, \mathbf{S}\,&\text{is the overlap matrix:}\;
S_{ij}=\langle\Phi_i|\Phi_j\rangle .
\end{align}
Here, $\mathbf{c}=(c_0,c_1,\dots,c_{d})^{\!\top}$ is the vector of expansion
coefficients, so the approximate ground state obtained from the subspace reads
\begin{equation}
  |\Psi_p\rangle=\sum_{i=0}^{d} c_i\,|\Phi_i\rangle .
\end{equation}
$E_p$ is the approximation of the ground state energy, which is always lower bounded by the true ground state energy.

\begin{bluebox}[label=box:CSM]{Classical Subspace Methods}
Classically, one projects the time-independent Schrödinger equation onto a
finite subspace \(V_n\) spanned by \(\{|v_\alpha\rangle\}_{\alpha=0}^{n-1}\),
yielding a generalized eigenvalue problem.  
Prominent approaches include:
\begin{itemize}
  \item \textbf{Krylov subspace methods}  
        \(K_n(H,|v_0\rangle)=\mathrm{span}\{|v_0\rangle,H|v_0\rangle,\dots,H^{n-1}|v_0\rangle\}\),
        delivering exponential convergence bounded by Kaniel–Paige–Saad
        inequalities \cite{saad1980rates,motta2024subspace}.
  \item \textbf{Lanczos algorithm}: orthonormalizes Krylov vectors and
        produces a tridiagonal Hamiltonian; numerical stability requires
        selective re-orthogonalization \cite{cullum2002lanczos, motta2024subspace}.
  \item \textbf{Davidson method}: augments the subspace with preconditioned
        residuals for rapid convergence
        \cite{DAVIDSON197587,motta2024subspace}.
  \item \textbf{Configuration-interaction (CI) hierarchy}:  CIS/CISD are variational yet
        non-size-consistent \cite{bender1969studies,motta2024subspace}.  
        Selected-CI algorithms (CIPSI, Heat-Bath, etc.) adaptively include only
        important determinants
        \cite{huron1973iterative,motta2024subspace}.
  \item \textbf{Equation Of Motion (EOM) techniques} for excited states rely on
        operator-commutator matrices and are size intensive
        \cite{bohm1951collective,motta2024subspace}.
\end{itemize}
\end{bluebox}
Selected Configuration Interaction (SCI) algorithms use subspaces spanned by an ortho-normal basis of Slater Determinants (SD), often called configurations, which simplifies $S_{ij}=\delta_{ij}$($\mathbf{S}=\mathbb{I}$), and allows for efficient estimation of the Hamiltonian terms $(\hat{H}_p)_{ij}$ through the Slater-Condon rules~\cite{tubman2020modern}. The configurations are adaptively selected, identifying the ones that make the largest contributions to the ground state.
In this way, subspace diagonalization methods approaches recover near‐exact energies while retaining only a tiny fraction of the full determinant set.

Building on the same principle, \emph{Quantum Subspace-diagonalization
Methods} (QSMs) transfer the costly basis-state enumeration to the quantum
processor.
A set of quantum states
\(\{\,|\Psi_i\rangle\}_{i=0}^{d}\) is prepared, the matrix elements \((\hat{H}_p)_{ij}=\langle\Psi_i|\hat H|\Psi_j\rangle\) and \(S_{ij}=\langle\Psi_i|\Psi_j\rangle\) are estimated on the quantum computer, and the generalized eigenvalue problem of Eq.~\eqref{eq:gen_eig} is then solved
classically.
QSMs thus combine shallow-circuit state preparation with classical linear algebra, inheriting the upper-bound property of classical subspace methods while targeting richer multi-state subspaces.   Box~\ref{box:QSM} surveys recent quantum subspace methodologies, for a comprehensive review of these methods we refer the reader to the recent review article Motta \emph{et al.}~\cite{motta2024subspace}.

\begin{bluebox}[label=box:QSM]{Quantum Subspace Methods}
\subsubsection*{Quantum Subspace Expansion (QSE)}
Applying $k$-body fermionic excitation operators to a quantum-prepared
reference state produces a subspace formally equivalent to multi-reference CISD,
with negligible extra circuit depth but substantial measurement overhead
\cite{mcclean2017hybrid,colless2018computation,huggins2020non,
       motta2020determining}.

\subsubsection*{Quantum Equation-of-Motion (qEOM)}
qEOM generalises EOM to arbitrary quantum states, measuring one- and
two-particle reduced density matrices; a self-consistent formulation restores size-intensivity
\cite{ollitrault2020quantum,asthana2023quantum}.

\subsubsection*{Krylov- and Time-Evolution–Based QSMs}
Chebyshev-Krylov methods require block-encoding \cite{kirby2023exact};  
Gaussian-power Krylov improves sampling \cite{zhang2024measurement};  
Quantum Filter Diagonalization (QFD) achieves rigorous error bounds via real-time
evolution \cite{epperly2022theory,kirby2401analysis,shen2023real};  
QLanczos/QITE replaces real-time with imaginary-time evolution while retaining
shallow circuits \cite{motta2020determining}.

\subsubsection*{Variational Subspace Hybrids}
Multistate contracted VQE jointly optimizes circuit parameters and linear
coefficients \cite{parrish2021analytical};  
non-orthogonal VQE leverages overlap-measuring protocols
\cite{huggins2020non,baek2023say}.

\subsubsection*{Eigenvector Continuation Methods}
Eigenvector continuation interpolates spectra across parameters  \cite{francis2022subspace,mejuto2023quantum}.

\end{bluebox}

Although, for a \(d\)-dimensional subspace, estimating all \(\mathcal{O}(d^{2})\) matrix elements \((\hat{H}_p)_{ij}\) and \(S_{ij}\) on quantum hardware requires deep circuits and controlled unitaries that rapidly exhausts a NISQ device’s coherence time and introduces significant errors.
Furthermore, because of the errors in the estimation of matrix elements, the strict variational upper-bound guarantee of the Ritz principle is no longer assured. The energies obtained from the noisy projection of the Hamiltonian \(\tilde{\mathbf H}_p\) and and overlap matrix \(\tilde{\mathbf S}\) can lie on either side of the exact ground state energy and are no longer a guaranteed upper bound.
To sidestep these issues, we turn to configuration-interaction–based quantum subspace methods that \emph{sample} determinants on the quantum device while evaluating all Hamiltonian matrix elements classically.  The next sections introduce the most recent NISQ-friendly approaches: Quantum-Selected Configuration Interaction (QSCI)~\cite{kanno2023quantum}, which presents the raw algorithm, and the state-of-the-art Sample-Based Quantum Diagonalization(SQD)~\cite{robledo2024chemistry}, which is built on QSCI with additional classical post-processing to further optimize the results.


\subsection{Quantum-Selected Configuration Interaction (QSCI)}
\label{sec:qsci}
The Quantum-Selected Configuration Interaction (QSCI) algorithm, first introduced by Kanno \emph{et al.}~\cite{kanno2023quantum}, is a hybrid quantum–classical method for estimating the ground-state energy of an electronic Hamiltonian $\hat{H}$. It utilizes a quantum processor to sample configurations (Slater Determinants) from an approximate eigenstate and a classical computer to perform exact diagonalization in a subspace defined by the most probable sampled configurations. The method guarantees variational energy estimates and is designed to be robust to quantum hardware noise.

\begin{enumerate}
    \item \textbf{State Preparation and Sampling:} Prepare a trial quantum state $\vert \psi_{\text{in}} \rangle$  (for example
        Hartree-Fock followed by a shallow UCC layer) that approximates the ground state of the Hamiltonian $\hat{H}$. This state may be prepared via multiple approaches such as VQE or time-evolutions algorithms~\cite{sugisaki2024hamiltonian, mikkelsen2412quantum}. We then measure $\vert \psi_{\text{in}} \rangle$ in the computational basis $N_{\text{shot}}$ times. Tabulate the frequencies of the observed bit-strings $\{ \vert x_i \rangle \}$, which correspond to Slater determinants.

    \item \textbf{Subspace Construction:} Select the $R$ most frequently measured bit-strings to form the subspace 
    \[
    \mathcal{S}_R = \left\{ \vert x_1 \rangle, \dots, \vert x_R \rangle \right\}.
    \]
    Multiple changes can be implemented in this phase, one can for example post-select measurement outcomes to enforce symmetries (e.g., fixed particle number or spin).

    \item \textbf{Hamiltonian Projection:} Compute the Hamiltonian matrix elements
    \begin{equation}
    (H_R)_{ij} = \langle x_i \vert \hat{H} \vert x_j \rangle, \quad \forall \vert x_i \rangle, \vert x_j \rangle \in \mathcal{S}_R,
    \end{equation}
    which is classically manageable using the Slater–Condon rules~\cite{tubman2020modern}.

    \item \textbf{Subspace Diagonalization:} Classically solve the eigenvalue problem $H_R c = E_R c$,
    where $E_R$ is the approximate ground-state energy, and $c$ is the coefficient vector of the eigenstate in the selected subspace, as shown in Equation~\eqref{eq:gen_eig}.

    \item \textbf{Wavefunction Reconstruction:} Construct the approximate ground-state wavefunction as
    \[
    \vert \psi_{\text{out}} \rangle = \sum_{i=1}^{R} c_i \vert x_i \rangle.
    \]
    This explicit wavefunction can be used to classically compute the expectation value of more observables enabling further analysis at no quantum cost.

\end{enumerate}

The QSCI energy $E_R$ satisfies the variational condition $E_R \geq E_{\text{exact}}$, and the quality of the result depends on the expressiveness of $\vert \psi_{\text{in}} \rangle$ and the size $R$ of the selected subspace. The method is particularly suitable for near-term quantum devices, as it avoids the error prone quantum expectation-value measurements and requires only projective measurements.

\subsection{Sample-Based Quantum Diagonalization (SQD)}
\label{subsec:SQD}
\begin{figure}
    \centering
    \includegraphics[width=0.8\linewidth]{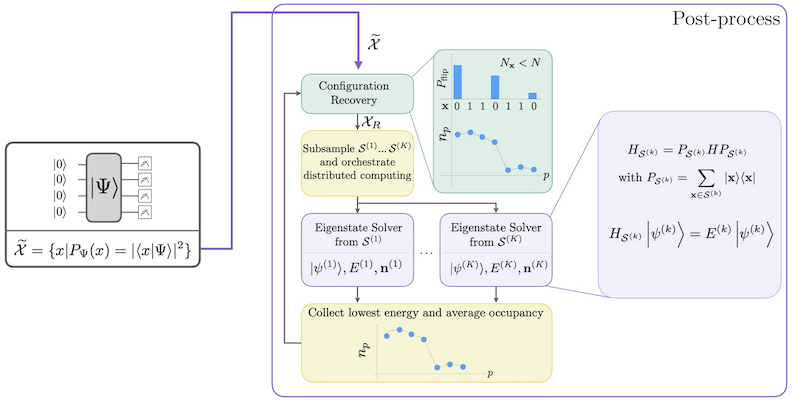}
    \caption{ An illustration of a Sample-Based Quantum Diagonalization workflow showcasing all the classical post-processing of the results measured on the quantum computer~\cite{qiskit-addon-sqd}.}
    \label{fig:sqd_workflow}
\end{figure}

Originally proposed in Ref.~\cite{robledo2024chemistry},
Sample-based Quantum Diagonalization (SQD) constitutes a hybrid
quantum–classical workflow built on QSCI, designed to increase the accuracy
and scaling of the method in order to better approximate molecular ground-state
energies. It uses a high-performance classical backend to post-process the
measured configurations and carries out independent diagonalizations in a set
of low-dimensional subspaces. 
Since its introduction, the method had seen various applications:  Simulation of \ce{N2} dissociation and [Fe-S] clusters~\cite{robledo2024chemistry}, supramolecular interactions~\cite{kaliakin2024accurate}, the singlet and triplet states of methylene~\cite{liepuoniute2024quantum}, and more realistic chemical molecules in a solvent~\cite{kaliakin2025implicit}.
The main steps of an SQD routine are summarized below and illustrated schematically in Fig.~\ref{fig:sqd_workflow}.

\begin{enumerate}
  \item \textbf{State preparation and sampling:}  
        Similarly to QSCI, we prepare a trial state $\vert \psi_{\text{in}} \rangle$ on the quantum device and measure it $m$ times in the computational basis to obtain the multiset $\mathcal{X}=\{x^{(1)},\dots,x^{(m)}\}$ of Slater determinants.

  \item \textbf{Configuration-pool and batch construction:}  
        Form an empirical probability distribution from $\mathcal{X}$ and draw
        $K$ disjoint batches
        $S^{(k)}=\{x^{(k)}_{1},\dots,x^{(k)}_{d}\}$,
        each of size $d$, with sampling weights proportional to their observed
        frequencies.  Determinants are filtered to respect conserved
        quantities (particle number, $S_z$, etc.).

  \item \textbf{Projected Hamiltonians and diagonalization:}  
        For every batch $S^{(k)}$ define the projector
        $\hat{P}_{S^{(k)}}=\sum_{x\in S^{(k)}}|x\rangle\langle x|$ and build
        the projected Hamiltonian
        \[
          H_{S^{(k)}}=\hat{P}_{S^{(k)}}\hat{H}\hat{P}_{S^{(k)}},
          \qquad
          \bigl[H_{S^{(k)}}\bigr]_{ij}=
          \langle x^{(k)}_{i}|\hat{H}|x^{(k)}_{j}\rangle .
        \]
        All matrix elements are evaluated classically via Slater–Condon
        rules, scaling as $\mathcal{O}(d^{2})$ per batch.  Each
        $H_{S^{(k)}}$ is then diagonalized with Davidson's iterative method to
        obtain its lowest eigenpair $E^{(k)},\,|\psi^{(k)}\rangle$.

  \item \textbf{Self consistent configuration recovery:} 
        Configurations with the wrong number of particles or wrong total spin are corrected probabilistically from the average occupation of each orbital in the ground states computed in the previous step $|\psi^{(k)}\rangle$.
        The loop batching $\rightarrow$ diagonalization $\rightarrow$ configuration recovery —continues until
        the spread $\max_k E^{(k)}-\min_k E^{(k)}$ falls below a user-defined
        tolerance.

  \item \textbf{Energy estimate and wavefunction reconstruction:}  
        The smallest batch energy,
        $E_{\text{SQD}}=\min_k E^{(k)}$, serves as the SQD ground-state
        estimate.  Its corresponding eigenvector
        $|\psi^{(k_{\min})}\rangle=\sum_{i=1}^{d}c^{(k_{\min})}_{i}
        |x^{(k_{\min})}_{i}\rangle$
        provides an explicit many-determinant wavefunction for further
        classical analysis at no additional
        quantum cost.
\end{enumerate}

Relative to QSCI, SQD replaces a single rank-\(R\) diagonalization with \(K\)
smaller problems of size \(d\), achieving an overall cost
\(\mathcal{O}(K\,d^{\alpha})\) with \(\alpha\approx1\!-\!2\) depending on the
iterative solver.  Quantum resources are identical to QSCI (state preparation
and projective measurements), but classical post-processing extracts more
information from the same shot budget, improving accuracy without additional
circuit depth.  Consequently, SQD offers a potentially scalable, NISQ-compatible pathway
to chemical accuracy for molecules beyond the reach of classical FCI.

The quantum sampling methods face a fundamental challenge that is discussed in chapter~\ref{chap3}.





\chapter{Challenges of near term algorithms}
\label{chap3}

\section{Measurement Overhead in Variational Algorithms}
A practical bottleneck for variational quantum algorithms is not the
preparation of the trial state itself but the \emph{sheer volume of
measurements} required to estimate the energy with chemical accuracy.
After mapping the electronic Hamiltonian onto a sum of Pauli strings
(Eq.~\eqref{eq:Pauli_sum}), each variational update demands that the
expectation value of every string be estimated to within a prescribed
statistical error.  Because the number of strings scales as
\(\mathcal{O}(n^{4})\) for an \(n\)-qubit chemistry Hamiltonian and shot noise
decays only as \(1/\sqrt{N_{\text{s}}}\), measurement overhead quickly
dominates the total runtime on current hardware.  The next subsection reviews
how individual Pauli expectations are obtained on a quantum processor and sets
the stage for grouping strategies.

\subsection{Estimating Pauli Expectation Values on a QPU}
\label{sec:measurement-pauli}

After mapping the electronic Hamiltonian to a qubit operator, Eq.~\eqref{eq:Pauli_sum}, the energy is the
weighted sum of Pauli‐string expectation values $E(\vtheta)= \sum_{i=1}^{P} w_{i}\matrixel{\psi(\vtheta)}{P_{i}}{\psi(\vtheta)}$.
The general method in which expectation values of Pauli strings are measured is described in what follows:
\paragraph{Direct $Z$‐basis measurement:}
For a single qubit measured in the computational basis, let
$p_{0}$ ($p_{1}$) denote the probability of obtaining outcome $0$ ($1$).
Because the eigenvalues of $Z$ are $\pm1$, the expectation value is
\begin{equation}
\langle Z\rangle
\;=\;
p_{0}-p_{1}
\;=\;
2p_{0}-1.
\label{eq:z-exp-from-probs}
\end{equation}
An unbiased empirical estimator obtained from $N_{s}$ shots is
\begin{equation}
\widehat{\langle Z\rangle}
\;=\;
\frac{n_{0}-n_{1}}{N_{s}},
\label{eq:z-estimator}
\end{equation}
where $n_{0}$ ($n_{1}$) is the observed number of $0$ ($1$) outcomes.

\paragraph{Basis rotations for non‐$Z$ observables:}
Quantum hardware natively measures only $Z$; to obtain
$\langle X\rangle$ or $\langle Y\rangle$ we apply single‐qubit
basis‐change gates \emph{before} the measurement:
\begin{equation}
X \;=\; HZH,
\qquad
Y \;=\; SHZH S^{\dagger},
\label{eq:basis-rotations}
\end{equation}
where $H$ is the Hadamard gate and $S=\sqrt{Z}$ is the phase gate.
After inserting these rotations, the measurement proceeds exactly as in
Eq.~\eqref{eq:z-exp-from-probs}.

\paragraph{Measuring an $n$‐qubit Pauli string:}
For a general Pauli string 
$P_{i}= \bigotimes_{q=1}^{n}\sigma^{(q)}_{i}$\footnote{the $(q)$ in the exponent is the number of the qubit on which the Pauli operator acts, and not an exponent.} we build a \emph{local}
unitary
\begin{equation}
R_{i}\;=\;\bigotimes_{q=1}^{n} R^{(q)}_{i},
\quad
R^{(q)}_{i}\;=\;
\begin{cases}
H, & \sigma^{(q)}_{i}=X,\\[4pt]
SH, & \sigma^{(q)}_{i}=Y,\\[4pt]
I, & \sigma^{(q)}_{i}=Z\text{ or }I.
\end{cases}
\label{eq:local-rotator}
\end{equation}
Because $R_{i}P_{i}R_{i}^{\dagger}=Z^{\otimes |\mathcal{S}_{i}|}$,
measuring the rotated state
$R_{i}^{\dagger}U(\vtheta)\ket{0}^{\otimes n}$
in the computational basis yields outcomes
$\mathbf{b}=(b_{1},\dots,b_{n})\in\{0,1\}^{n}$, from which we compute the empirically estimated expectation value (i.e average over $N_s$ shots)
\begin{equation}
\widehat{\langle P_{i}\rangle}
\;=\;
\frac{1}{N_{s}}\sum_{k=1}^{N_{s}}
\;\prod_{q\in\mathcal{S}_{i}} (-1)^{\,b_{k,q}} .
\label{eq:pauli-string-estimator}
\end{equation}
Here $\mathcal{S}_{i}$ is the set of qubits on which $P_{i}$ acts
non‐trivially and $b_{k,q}$ is the $q$‐th bit of the $k$‐th sample.

\subsection{Shot Allocation and Pauli-String Grouping}
\label{sec:shots_and_grouping}

When an observable is estimated by repeated projective measurements, the
standard error obeys
\(
  \epsilon=\sigma/\sqrt{N_s},
\)
where \(\sigma\) is the true standard deviation of the measured random
variable and \(N_s\) is the total number of circuit executions
(\emph{shots}).  Consequently the shot count required to reach a target
precision scales as \(\mathcal{O}(1/\epsilon^{2})\).

If the Hamiltonian has already been decomposed into
\(
  H=\sum_{i=1}^{\mathcal{P}} w_{i}P_{i},
\)
and the available shots are distributed \emph{optimally} among the
individual Pauli terms so as to minimise the total variance, the number
of shots satisfies the upper bound
\begin{equation}
  N_s \;\le\;
  \Bigl(\tfrac{\sum_{i} |w_{i}|}{\epsilon}\Bigr)^{\!2}.
  \label{eq:global_shot_bound}
\end{equation}
For electronic-structure Hamiltonians the term count grows as
\(\mathcal{P}\sim\mathcal{O}(n^{4})\) for \(n\) qubits, implying
\(
  N_s = \mathcal{O}\!\bigl(n^{4}/\epsilon^{2}\bigr)
\)
in the worst case.

If each Pauli string is measured separately, the total mean-squared
error of the energy estimator is
\begin{equation}
  \epsilon^{2}
  \;=\;
  \sum_{i=1}^{\mathcal{P}}
  w_{i}^{\,2}\,
  \frac{\mathbb{V}(P_{i})}{N_{i}}
  \;=\;
  \sum_{i=1}^{\mathcal{P}}
  w_{i}^{\,2}\,
  \frac{1-|\langle P_{i}\rangle|^{2}}{N_{i}},
  \label{eq:mse_independent}
\end{equation}
with \(N_{i}\) shots assigned to \(P_{i}\) and \(\sum_{i} N_{i}=N_{s}\).
Because every Pauli operator squares to the identity, $\mathbb{V}(P_{i}) = 1-|\langle P_{i}\rangle|^{2}\le 1$ for any state.

When working with a fixed budget of shots in total $N_s$, Rubin \emph{et al.}~\cite{rubin2018application} showed that the variance in
Eq.~\eqref{eq:mse_independent} is minimized by the allocation
\(N_{i}\propto |w_{i}|\sqrt{\mathbb{V}(P_{i})}\).
Arrasmith \textit{et al.}~\cite{arrasmith2020operator} further noted that one
can use the simpler heuristic \(N_{i}\propto |w_{i}|\), avoiding the need
to estimate $\mathbb{V}(P_{i})$ itself.
\paragraph{Grouping measurements: }
Measuring every term separately soon becomes impractical.  A more
efficient strategy is to partition the Pauli operators into mutually
commuting subsets.  If a collection \(\{P_{i}\}_{i\in\alpha}\) forms an
Abelian subgroup, there exists a unitary \(U_{\alpha}\) such that
\(U_{\alpha}^{\dagger}P_{i}U_{\alpha}\) is diagonal in the computational
basis for all \(i\in\alpha\).  Executing
\(U_{\alpha}^{\dagger}\) prior to measurement therefore yields
simultaneous estimates of every generator in the set, and the remaining
member operators can be reconstructed from the same bit-strings.\footnote{For
a survey of grouping techniques, see Chapter~5 of
Tilly \textit{et al.}~\cite{tilly2022variational}.}

Multiple strategies exist that tackle this challenge. For a chemistry focused analysis of the measurement requirements for variational quantum algorithms, we refer the reader to the recent work by Gonthier \emph{et al.}~\cite{gonthier2022measurements}.

\section{Barren plateaus in VQE}

In VQE, our aim is to minimize the energy $E(\vtheta)$, defined in Eq.~\eqref{eq:VQE_energy}  to approach the ground state energy $E_0$ of $H$. However, as the system size increases, VQE optimizations suffer from a critical scalability bottleneck known as the Barren Plateau (BP) phenomenon~\cite{mcclean2018barren, cerezo2021cost}. 
In this section, we will elaborate on this issue by presenting its theoretical formulation, identifying its origin in the geometry of high-dimensional Hilbert spaces, and discussing how the structure of the ansatz, particularly its dynamical Lie algebra, plays a crucial role in determining whether or not the landscape exhibits exponentially vanishing gradients.

\paragraph{Notation:}
Throughout this section, we denote the variational quantum state as $\ket{\psi(\vtheta)} = U(\vtheta) \ket{\psi_0}$, where $U(\vtheta)$ is a parameterized quantum circuit with parameters $\vtheta \in \mathbb{R}^d$, and $\ket{\psi_0}$ is a fixed reference (initial) state ($\rho_0=\ket{\psi_0}\bra{\psi_0}$).

The cost function corresponding to the energy expectation value with respect to a target Hamiltonian $H$ is denoted by
\begin{equation}
    E(\vtheta) = C_{\vtheta}(\rho_0, H) = \mathrm{Tr}[H\,U(\vtheta)\,\rho_0\,U(\vtheta)^\dagger].
\end{equation}

\begin{figure}[h]
    \centering
    \includegraphics[width=\textwidth]{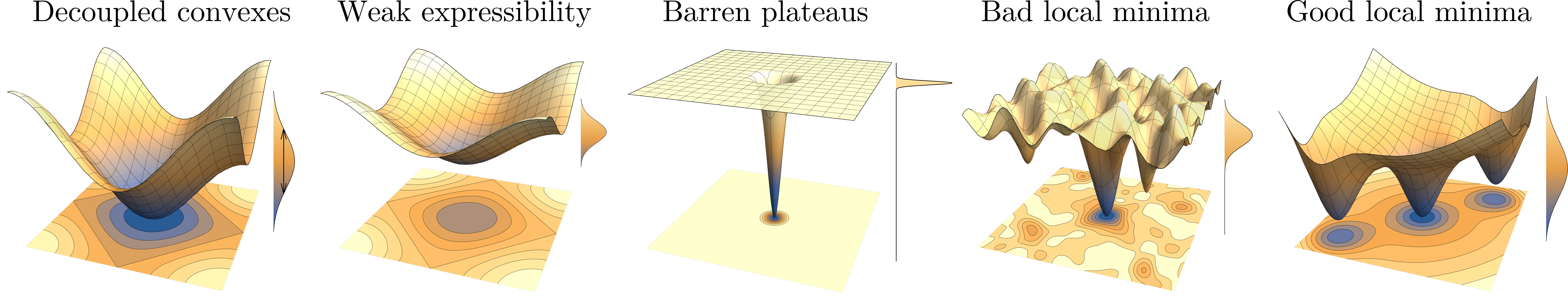}
    \caption{Plots of typical training landscapes, with the density of states cost value shown on the right side. Although the parameter space is inherently high-dimensional, it is presented in 2D here for ease of visualization. Figure adapted from Zhang \emph{et al.} \cite{zhang2406predicting}.}
    \label{fig:bp}
\end{figure}

\paragraph{Definition:} 
A \emph{barren plateau}~\cite{larocca2025barren, mcclean2018barren} refers to the phenomenon where landscape of the cost function becomes exponentially flat with respect to the number of qubits $n$. More precisely, the variance of the cost function $C_{\vtheta}(\rho, H)$ or its gradient component $\nabla_{\theta_\mu} C_{\vtheta}(\rho, H)$ scales as
\begin{equation}
\label{eq:variances}
\mathrm{Var}_{\vtheta}[C_{\vtheta}(\rho_0, H)] \quad \text{or} \quad \mathrm{Var}_{\vtheta}\left[\frac{\partial C_{\vtheta}(\rho_0, H)}{\partial \theta_{\mu}}\right] \in \mathcal{O}\left( \frac{1}{b^n} \right),
\end{equation}
for some $b > 1$ and for at least one parameter $\theta_{\mu} \in \vtheta$, corresponding to the case of \emph{probabilistic} concentration.
In contrast, a \emph{deterministic} barren plateau arises when the cost function becomes exponentially concentrated around a constant value $C_0$ that is independent of $\vtheta$, such that
\begin{align}
    |C_{\vtheta}(\rho_0, H) - C_0| \in  \mathcal{O}\left( \frac{1}{b^n} \right)
\end{align}
for some $b > 1$.
In such cases, optimization becomes exponentially hard: the gradients vanish in most directions, rendering gradient-based and even gradient-free optimizers ineffective.

\subsection{Origins of barren plateaus}
We will consider a Lie Algebraic Theory~\cite{ragone2024lie} that allows us to rigorously define the variance from Eq.~\eqref{eq:variances}, allowing us to explain the sources of barren plateaus.
\paragraph{Dynamical Lie Algebras (DLAs):}
The concept of Lie subalgebras $\mathfrak{g} \subseteq \mathfrak{su}(\mathcal{H})$ plays a pivotal role in the analysis of barren plateaus. Here, $\mathfrak{su}(\mathcal{H})$ denotes the space of traceless anti-Hermitian operators acting on the Hilbert space $\mathcal{H}$. Any subset $\mathfrak{g} \subset \mathfrak{su}(\mathcal{H})$ that is closed under the Lie bracket (i.e., commutators) forms a Lie subalgebra. For a unitary, noiseless parameterized quantum circuit (ansatz) of the form
\[
U(\vtheta) = \prod_{\ell} e^{-i \theta_\ell H_\ell},
\]
the associated DLA~\cite{wiersema2024classification,zeier2011symmetry,larocca2022diagnosing} is defined as
\[
\mathfrak{g} = \langle \{ i H_\ell \} \rangle_{\mathrm{Lie}} \subseteq \mathfrak{su}(\mathcal{H}),
\]
where the notation $\langle S \rangle_{\mathrm{Lie}}$ denotes the Lie closure of the set $S$, i.e., the smallest real vector space that contains $S$ and is closed under repeated commutators.

The dynamical Lie algebra $\mathfrak{g}$ is a central object in understanding the expressive power of variational quantum circuits. The corresponding Lie group $G = e^{\mathfrak{g}}$ consists of all unitaries that can be generated by exponentiating elements of $\mathfrak{g}$. Therefore, $G$ characterizes the set of all unitary operations that are accessible via combinations of the circuit’s generating gates $H_\ell$ for some choice of parameters $\vtheta$ and circuit depth $L$.
This connection can be explicitly established via the Baker–Campbell–Hausdorff (BCH) formula, which expresses a product of exponentials of operators as the exponential of a linear combination of those operators and their nested commutators. Thus, the reachable unitary operations lie within the Lie group $G = e^{\mathfrak{g}}$, and the dynamical Lie algebra $\mathfrak{g}$ governs the structure of the parameter space explored by the circuit.

Following Ragone \emph{et al.}~\cite{ragone2024lie}, when $\mathfrak{g}$ is simple, the variance of the cost function is given by:
\begin{align}
\label{eq:lie_variance}
    \mathrm{Var}[C_{\vtheta}(\rho_0, H)] = \frac{\mathcal{P}_{\mathfrak{g}}(\rho_0)\mathcal{P}_{\mathfrak{g}}(H)}{\dim(\mathfrak{g})}
\end{align}
where $\mathcal{P}_{\mathfrak{g}}(\rho_0)$ is the $\mathfrak{g}$-purity of $\rho_0$~\cite{barnum2004subsystem} defined as $\mathcal{P}_{\mathfrak{g}}(O)= \sum_{H_\ell \in \mathfrak{g}} |Tr[H_\ell O]|^2$ . $\dim(\mathfrak{g})$ is the dimension of the DLA underlining the Parametrized Quantum Circuit (PQC), it is a measure of the expressiveness of the PQC. When $\dim(\mathfrak{g})= 4^n -1$ we say that the PQC is maximally expressive.

\begin{figure}
    \centering
    \includegraphics[width=0.99\linewidth]{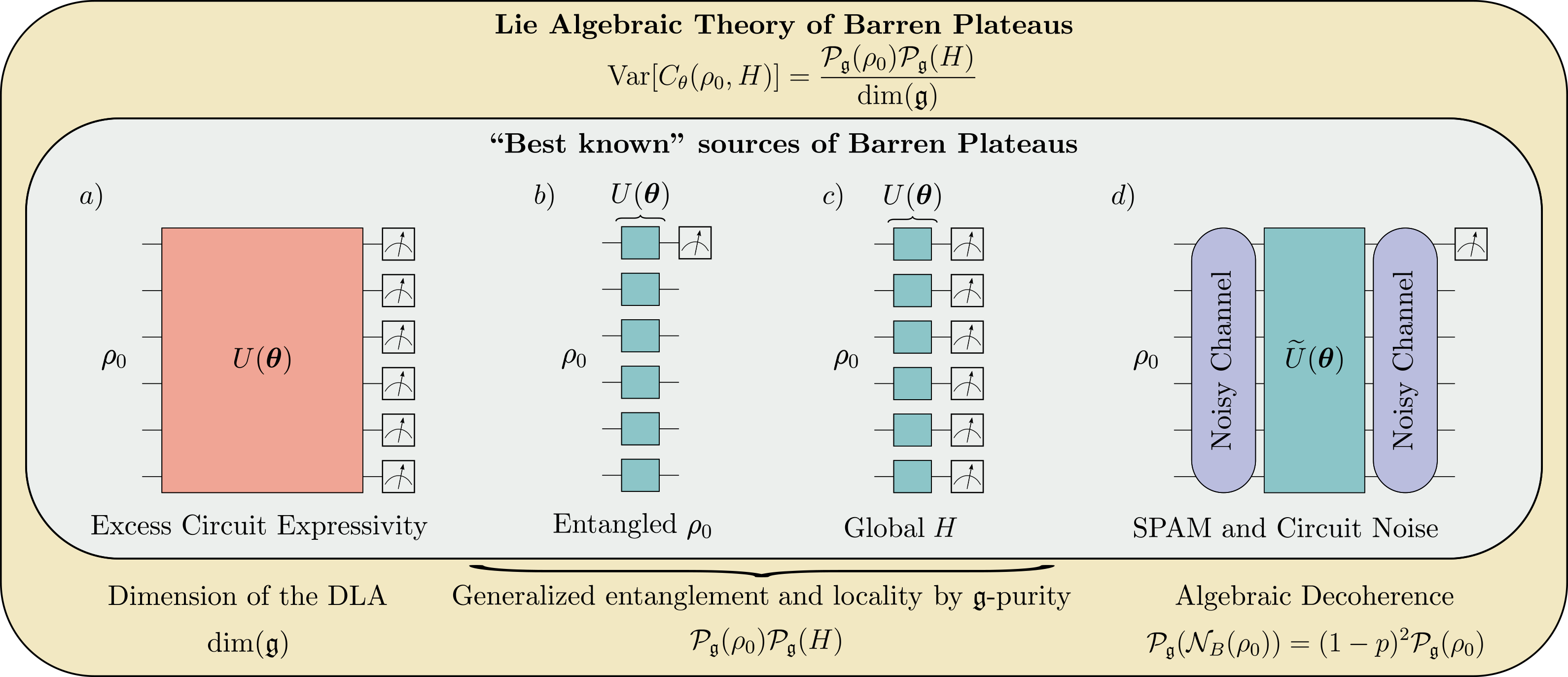}
    \caption{The known sources of barren plateaus in the Lie Algebraic Theory Formulation, Figure adapted from Ragone \emph{et al.}~\cite{ragone2024lie}.}
    \label{fig:BP_lie}
\end{figure}
Following Eq.~\eqref{eq:lie_variance}, Ragone \emph{et al.} identify the following sources of barren plateaus~\cite{ragone2024lie}, which are illustrated in Fig.~\ref{fig:BP_lie}
\begin{enumerate}
    \item \textbf{Expressiveness of the Ansatz:} when the ansatz is over-expressive, $\dim(\mathfrak{g}) \in \mathcal{O}(b^n)$ corresponding to $ \mathrm{Var}[C_{\vtheta}(\rho_0, H)] \in\mathcal{O}(\frac{1}{b^n})$. As a result, expectation values become highly concentrated around their mean~\cite{mcclean2018barren}.
    \item \textbf{Initial State $\rho_0$:} We consider the $\mathfrak{g}$-purity of the initial state $\mathcal{P}_{\mathfrak{g}}(\rho_0)$, this quantity has been studied in literature and is known to be a measure of generalized entanglement~\cite{barnum2004subsystem}. This quantity exhibits exponential decay $\mathcal{P}_{\mathfrak{g}}(\rho_0) \in \mathcal{O}(\frac{1}{b^n}) $ when $\rho_0 $ is a high generalized entangled state~\cite{grant2019initialization}.
    \item \textbf{Measured Observable $H$:} We consider the $\mathfrak{g}$-purity of the Hamiltonian $\mathcal{P}_{\mathfrak{g}}(H)$, we carry over the definition of generalized entanglement from $\mathfrak{g}$-purity to define a generalized notion of locality. We call an operator $O$ generalized-local if it belongs to the preferred
    subspace of observables given by $\mathfrak{g}$, in wchich case, $O_\mathfrak{g} = O$. On the other hand, we will call it (fully) generalized nonlocal if $O_\mathfrak{g} = 0$. When the Hamiltonian $H$ is generalized non-local (global), $\mathcal{P}_{\mathfrak{g}}(H)\in \mathcal{O}(\frac{1}{b^n})$, rendering the variance exponentially concentrated around its mean value~\cite{holmes2022connecting}.
    \item \textbf{Depolarizing Noise~\cite{wang2021noise}:} We conclude by examining the impact of noise and errors of NISQ era devices on the variance scaling. We only consider the case of depolarizing noise, which alters the initial state from $\rho_0$ to a noisy version $\mathcal{N}_B(\rho_0)$. According to Eq.~\eqref{eq:lie_variance}, a decrease in the $\mathfrak{g}$-purity of the state leads to a corresponding decrease in the variance. This behavior arises under global depolarizing noise, modeled as $\mathcal{N}_B(\rho_0) = (1 - p)\rho_0 + p \mathbb{I}/2^n$, which transforms the purity as follows: $\mathcal{P}_{\mathfrak{g}}(\rho_0) \mapsto \mathcal{P}_{\mathfrak{g}}(\mathcal{N}_B(\rho_0)) = (1 - p)^2 \mathcal{P}_{\mathfrak{g}}(\rho_0)$. It is worth noting that sources of noise besides depolarization can be modeled and have also been shown to cause BPs~\cite{ragone2024lie}.

\end{enumerate}



\paragraph{Mitigation Strategies:} Various approaches have been proposed to mitigate BPs in VQE, including:
\begin{itemize}
    \item \emph{Ansatz design with small dynamical Lie algebras}~\cite{larocca2023theory}: For example, truncated UCC ansatzes preserve particle number and restrict the search space.
    \item \emph{Shallow circuits or local observables}~\cite{cerezo2021cost, uvarov2021barren}: Local energy terms can exhibit polynomial variance scaling even when the global cost landscape is flat.
    \item \emph{Informed parameter initialization}~\cite{grant2019initialization}: Starting near the ground state can increase gradient magnitude at initialization and help avoid flat regions.
\end{itemize}

Despite these strategies, BPs remain a persistent obstacle. The trade-off between trainability and expressivity is particularly pronounced in quantum chemistry, where ans{\"a}tze must capture subtle electron correlation effects but remain shallow and structured enough to be trainable on noisy hardware. For a thorough review of barren plateaus in variational quantum computing, we refer the reader to the recent review by Larocca \emph{et al.}~\cite{larocca2025barren}.

The barren plateau problem emphasizes the need for careful ansatz engineering and initialization protocols in VQE. While promising mitigation strategies exist, none offer a full resolution at scale. Thus, BPs are widely recognized as a fundamental limitation on the scalability of VQE for quantum chemistry, and understanding their emergence remains an active area of research~\cite{cerezo2022challenges}.

\section{Bottlenecks in Sampling-Based Methods}
\label{sec:sampling_challenges}
Since the introduction of Sample-based Quantum Diagonalization (SQD)~\cite{robledo2024chemistry} and showcasing its ability to tackle large systems with current hardware, the sampling-driven approaches gained attention as the potential candidate to achieve quantum advantage. However, there exists a bottleneck that is yet to be overcome in these methods. The problem was first flagged by Reinholdt \emph{et al.}~\cite{reinholdt2025exposing} where they numerically investigated the effectiveness of QSCI, the backbone of SQD. The process of sampling is inefficient: far fewer unique configurations were observed compared to the number of projective measurements performed (shots). The configurations with high a probability in the ground state keep being resampled multiple times instead of sampling new, unseen ones. The more configurations are discovered, the harder it is to discover new ones. In this section, we investigate this problem by linking it to the famous coupon collector problem~\cite{boneh1997coupon} from statistics. We establish a  theoretical formula for the average number of measurements required. We follow by introducing an approximation to efficiently compute a lower bound for that formula.

\subsection{Idealized sampling assumption}
Sampling-driven algorithms such as QSCI and SQD, in the ideal case, assume having an efficient-to-prepare \emph{exact} many-electron ground state:
\begin{equation}
   \bigl|\psi_{\mathrm{GS}}\bigr\rangle
      =\sum_{i=1}^{\text{dim}(\mathcal{A})} c_i\,\bigl|\phi_i\bigr\rangle,
   \qquad
   \mathcal{A}=\mathrm{span}\{\lvert\phi_i\rangle\}
   \subset \mathcal{H},
   \label{eq:true_ground_state}
\end{equation}
that when repeatedly measured will reveal the set of Slater determinants (Configurations/bit-strings)
\(\mathcal{S} \equiv\{\lvert\phi_i\rangle\}\).
We consider $\mathcal{A}$ to be the minimal subspace needed to represent the ground state. We call the set $\{\lvert\phi_i\rangle\}$ the support of the ground state.
Crucially, the \emph{amplitudes} \(c_i\) need not be learned; one must
merely \emph{discover} which determinants have a non-zero contribution to the ground state to get the exact energy of the ground state.

The practical cost of this discovery task depends not only on the size
\(m=\text{dim}(\mathcal{A}) = |\mathcal{S}|\) but also on the shape of the probability vector
\(\mathbf{p}=(p_1,\dots,p_m)\) with components \(p_i=\lvert c_i\rvert^2\).

\subsection{Connection to the coupon-collector problem}
Sampling until every determinant appears at least once is isomorphic to
the classical \emph{coupon-collector problem with unequal
probabilities}~\cite{flajolet1992birthday}.  The expected (average) number of
projective measurements (shots) until all the determinants are discovered is
\begin{equation}
   \tilde{M}=\sum_{r=0}^{m-1}(-1)^{\,m-1-r}
   \!\!\sum_{\substack{J\subseteq\mathcal{S}\\|J|=r}}
   \frac{1}{1-\sum_{i\in J} p_i},
   \label{eq:coupon_collector}
\end{equation}
a quantity dominated by the small denominators (rare configurations). 

For example, if we consider the set of configurations $\{|\alpha\rangle,|\beta\rangle,|\gamma\rangle\}$ with probability vector $\mathbf{p}_{\text{example}}=(a,b,c)$, Then the average number of measurements until all configurations are discovered is give by:
\begin{equation}
    \tilde{M}_{\text{example}}=1-\frac{1}{1-a}-\frac{1}{1-b}-\frac{1}{1-c}+\frac{1}{1-a-b}+\frac{1}{1-a-c}+\frac{1}{1-b-c}
    \label{eq:example}
\end{equation}

\paragraph{Uniform amplitudes (best case):}
If the ground-state amplitudes are perfectly uniform, $p_i=1/m\;\forall i$, the inner sums telescope to the well-known expression:
\begin{equation}
   \tilde{M}=m\,H_m
       =m\!\left(1+\tfrac12+\cdots+\tfrac1m\right)
       \approx m\ln m,
   \label{eq:uniform_cost}
\end{equation}
where \(H_m\) is the \(m\)-th harmonic number.  The \(m\ln m\) scaling
is the theoretical optimum for any sampling-based discovery protocol.

\paragraph{Skewed amplitudes:}
When one determinant carries almost all the weight, computing the exact shot count from Eq.~\eqref{eq:coupon_collector} becomes impractical:
the outer sum runs over every subset of $\mathcal{S}$, i.e.\ $\mathcal{O}(\binom{m}{r})$ 
unique terms that need to be computed separately because of their unique probabilities. For example, when $m=60$ there exists $\approx 6\times10^{16}$ terms that need to be computed separately, hence, a more computationally tractable estimate is needed. It is worth noting that the formula in Eq.~\eqref{eq:coupon_collector} is numerically unstable for large $m$ and skewed amplitudes. 

\begin{figure}[htbp]
  \centering
  \begin{subfigure}[t]{0.49\textwidth}
    \centering
    \includegraphics[width=\linewidth]{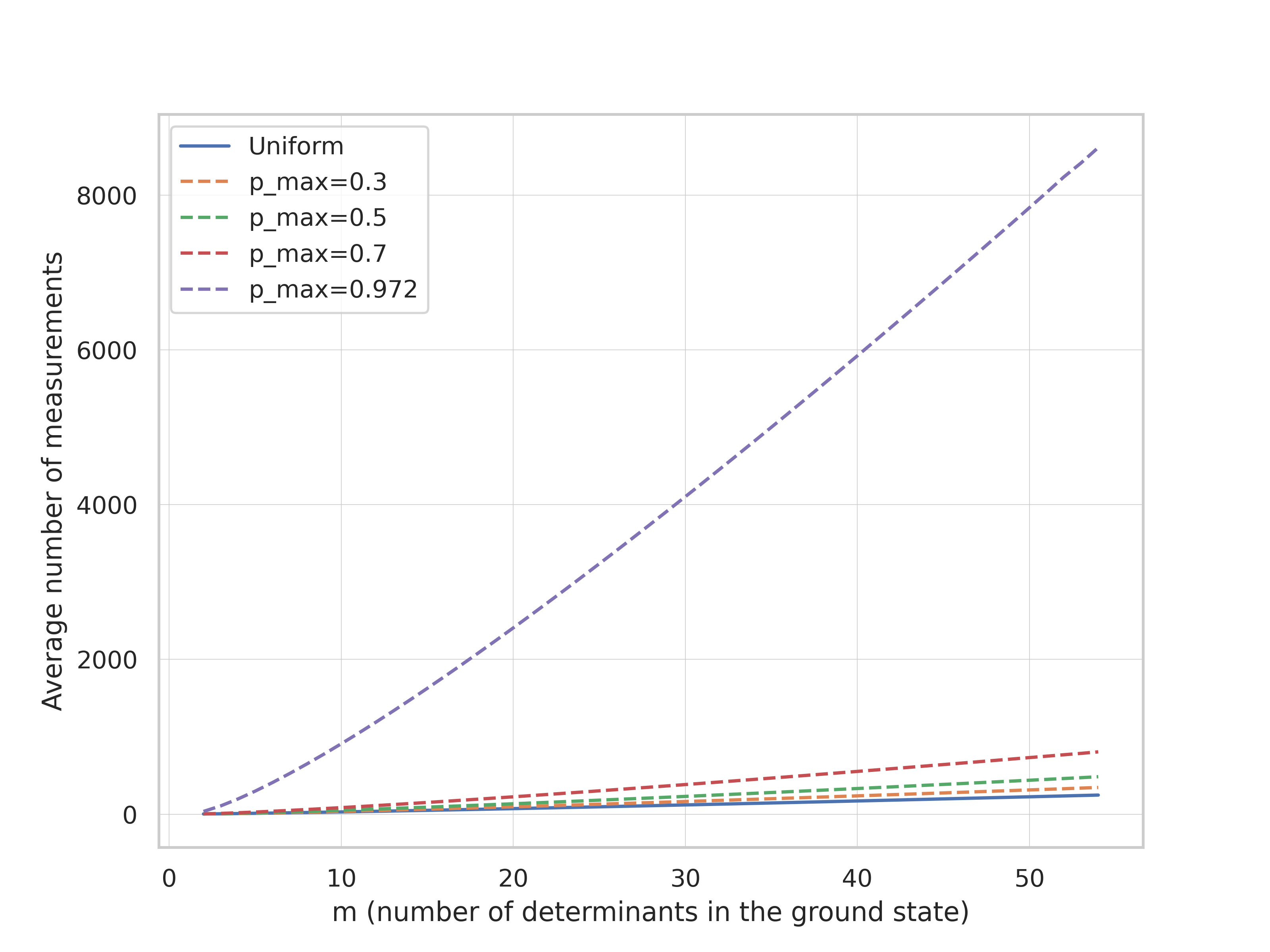}
    \caption{Expected number of measurements as a function of the number of determinants \(m\) in the ground state. Distributions with large \(p_{\max}\) result in significantly higher sampling costs.}
    \label{fig:subfig1}
  \end{subfigure}
  \hfill
  \begin{subfigure}[t]{0.49\textwidth}
    \centering
    \includegraphics[width=\linewidth]{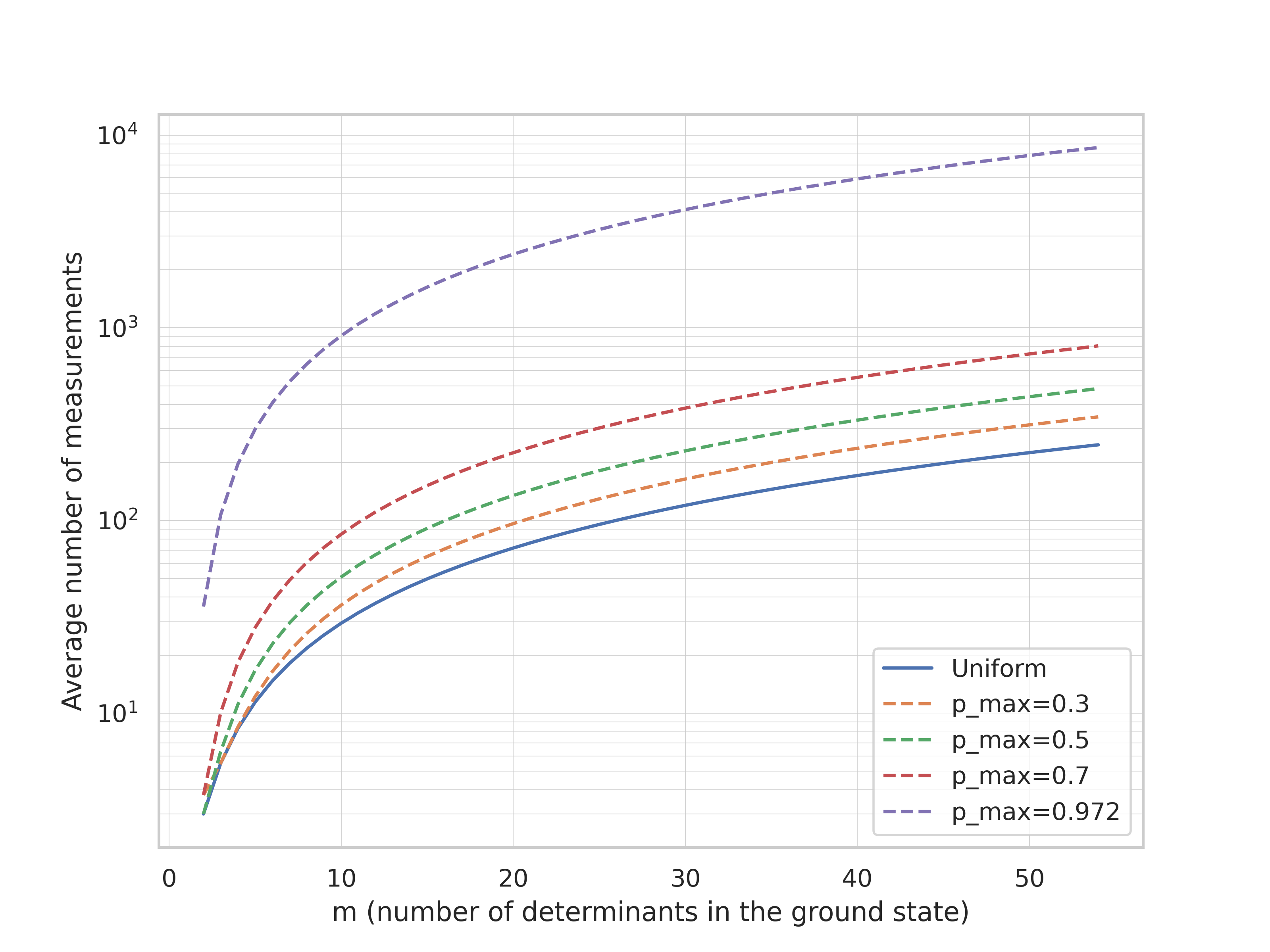}
    \caption{Same data shown on a log scale to highlight the scaling of measurement cost with \(m\) in skewed distributions.}
    \label{fig:subfig2}
  \end{subfigure}

  \caption{Coupon-collector cost for discovering all determinants in the ground state under various amplitude distributions. Both panels use the lower-bound estimate of Eq.~\eqref{eq:skew_bound}.}
  \label{fig:coupon_collector}
\end{figure}

\medskip
\noindent\textbf{Lower-bound approximation:}\\
Let
\[
   p_{\max} \;=\;\max_i p_i .
\]
We build a new, easier to work with, probability vector
\[
   q_{1} \;=\;p_{\max}, \qquad
   q_i \;=\;\frac{1-p_{\max}}{m-1}\;= \tilde{q}\;\;(i\ge2),
\]
which keeps the largest weight unchanged but spreads the rest out
equally.  
Because each $q_i$ for $i\ge2$ is the same, the waiting time for the
$\mathbf{q}$ distribution is a \emph{lower bound} for the true waiting
time. 
Plugging the components of $\mathbf{q}$ into the coupon-collector formula splits the second sum into two terms that are easy to compute:
\begin{equation}
    \sum_{\substack{J\subseteq\mathcal{S}\\|J|=r}}
   \frac{1}{1-\sum_{i\in J} q_i}= \binom{m-1}{r-1} \frac{1}{1-q_1-(r-1)\tilde{q}}+ \binom{m-1}{r} \frac{1}{1-(r)\tilde{q}}
\end{equation}
and since $q_1+(m-1)\tilde{q}=1$, the first term can be written as:

\begin{equation}
    \sum_{\substack{J\subseteq\mathcal{}S\\|J|=r}}
   \frac{1}{1-\sum_{i\in J} q_i}= \binom{m-1}{r-1} \frac{1}{(m-r)\tilde{q}}+ \binom{m-1}{r} \frac{1}{1-(r)\tilde{q}}
\end{equation}
and gives a new, easier to compute, expression:
\begin{equation}
   \tilde{M}=\sum_{r=0}^{m-1}(-1)^{\,m-1-r}\left[ \binom{m-1}{r-1} \frac{1}{(m-r)\tilde{q}}+ \binom{m-1}{r} \frac{1}{1-(r)\tilde{q}}\right]
   \label{eq:skew_bound}
\end{equation}
which we use to efficiently compute a lower bound to Eq. \eqref{eq:coupon_collector} with large systems that we plot in Fig.~\ref{fig:coupon_collector}.
For the example in Eq.~\eqref{eq:example}, assuming $a>b>c$, then $\tilde{q}=\frac{1-a}{2}$ $\mathbf{q}_{\text{example}}=(a,\tilde{q},\tilde{q})$
and the lower bound becomes:
\begin{equation*}
   \tilde{M}_{\text{example}}=1-\frac{1}{2\tilde{q}}-\frac{2}{1-\tilde{q}}+\frac{2}{\tilde{q}}+\frac{1}{1-2\tilde{q}} 
\end{equation*}
This approximation reduces the number of fraction terms that need to be computed separately and stored from $\sum_{r=0}^{m-1} \binom{m}{r}$  to $2 \times(m-1)$ while maintaining a lower bound that is good enough to show the scaling cost with skewed amplitudes.



\medskip
This analysis demonstrates that the core difficulty in sampling-based quantum diagonalization methods lies not in the estimation step but in discovering the correct support of the ground state. When the ground state is heavily dominated by a single determinant, as is common in weakly to moderately correlated molecules that are dominated by the Hartree–Fock configuration, the measurement process tends to resample this dominant configuration, delaying the appearance of rare (yet chemically essential) determinants. 
The left panel of Fig.~\ref{fig:coupon_collector} confirms this: as the number of relevant determinants \(m\) increases, the expected number of measurements remains manageable only for uniform or mildly skewed amplitudes. However, when \(p_{\max}\!\to\!1\), the cost explodes rapidly, and by orders of magnitude, even at moderate \(m\approx55\), which is the number of important configurations in the water-molecule ground state considered in Chapter~\ref{chap4}. This divergence is even clearer in the log-scale right panel. For instance, with \(p_{\max}=0.972\), which we later encounter for \ce{H2O}, the \emph{lower bound} on discovery cost already exceeds 8 000 measurements at \(m=50\), nearly two orders of magnitude worse than the uniform case. Extrapolating to larger systems, Reinholdt~\cite{reinholdt2025exposing} estimates that exhaustive discovery in \ce{N2} (6-31G) would require \(\sim10^{14}\) measurements; at a few milliseconds per shot, that translates to \(\approx10^4\) years of runtime. The bottleneck is not merely theoretical: Chapter~\ref{chap4} shows the same slowdown when sampling from the approximate \ce{H2O} ground state even in the ideal simulator. Consequently, the fundamental challenge for sampling-based methods is not the classical diagonalization step but discovering the relevant configurations within tractable shot budgets.

\chapter{Computational Setup and Results}
\label{chap4}
\section{Computational Setup}

\subsection{Molecule}
\label{sec:h2o_setup}

\begin{figure}[h]
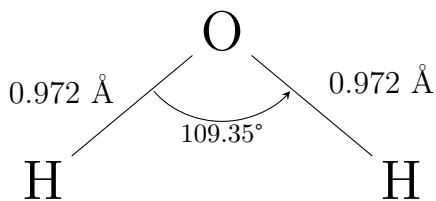

  \centering
  \vspace{1cm}

  \setchemfig{
    bond offset = 4pt,
    atom sep    = 50pt,
    atom style  = {scale=1.75},
    angle increment = 1 
  }

  \chemfig{@{hL}H-[:40]@{O}O-[:-40]@{hR}H}

  \namebond{hL}{O}{above}{$0.972$ \AA\quad\quad\quad\quad\quad\quad}
  \namebond{O}{hR}{above}{\quad\quad\quad\quad $0.972$ \AA}

  \arclabel{1.2cm}{hL}{O}{hR}{\footnotesize 109.35\textdegree}

  \vspace{0.5cm}
  \caption{Equilibrium gas-phase geometry of \ce{H2O} in the STO-3G basis~\cite{NIST_CCCBDB_2022}.}
  \label{fig:h2o_geometry}
\end{figure}

All quantum-chemistry simulations were performed on the water molecule in its gas-phase equilibrium geometry~\footnote{
\(O(0.0000,0.0000,0.1125)\), \(H(0.0000,0.7938,-0.4500)\), and \(H(0.0000,-0.7938,-0.4500)\)} as illustrated in Figure \ref{fig:h2o_geometry}, using the minimal STO-3G basis set~\cite{NIST_CCCBDB_2022}.
This initially gives $10$ electrons in an active space of seven spatial orbitals (fourteen spin-orbitals). We perform an active space reduction by freezing the \(1s\) orbital of the Oxygen atom only leaving $8$ electrons in six spatial orbitals (twelve spin-orbitals) that are mapped one-to-one onto 12 qubits via the Jordan–Wigner mapping, with no symmetry based qubit tapering. The ground state of this molecule in this setup is a closed-shell singlet state ($S=0,\, S_z=0$) with a full-CI benchmark energy of \(E_\text{exact}=-75.01259\,\text{Ha}\).

The \ce{H2O} molecule in this active space is strongly correlated: its
restricted Hartree-Fock reference energy lies at
\(E_{\mathrm{HF}}\approx -74.9619\;\text{Ha}\),
i.e.\ about \(\lvert\Delta E_{\mathrm{HF}}\rvert \simeq
50\,\text{mHa}\) (milliHartree) above the full‐CI benchmark.
Because chemical accuracy is customarily set at
\(\pm 1.6\,\text{mHa}\), the gap is more than an order of magnitude
larger, leaving ample room for post‐HF correlation methods and the estimators studied here to demonstrate clear improvements.
The one- and two-electron integrals for the \ce{H2O} molecule in this geometry and active space were obtained with the python package PySCF~\cite{pyscf}.

\subsection{Ground State Preparation}
\label{sec:ground_state_prep}
Starting from the restricted Hartree–Fock reference state (closed-shell singlet, \(S=0\)), a UCCSD
ansatz was used to prepare the approximate ground state and its amplitudes were optimized on an ideal state-vector simulator using the Excitation Solver optimizer, introduced by Jager \emph{et al.}~\cite{jager2024fast}, that is tailored to physically motivated excitations based ans{\"a}tze. The optimization was terminated just after 3 iterations when the estimated energy was within $0.2$ mHa of the exact energy \(E_{\text{exact}}= -75.01259\,\text{Ha}\).
The converged variational state is considered as the approximate ground state $|\psi_\text{in}\rangle$ which we will sample from and evaluate its energy with the two estimators introduced in Chapter~\ref{chap2}: Quantum Subspace Configuration Interaction (QSCI), and Sample-based Quantum Diagonalization (SQD), each under varying total shot budgets to quantify their performance.

We test the different estimators under different conditions: First on an ideal simulator to gauge their theoretical performance,  then in a noisy simulator to test their noise-resilience, and finally on the real quantum computer \textsc{IBM Brisbane}.

\subsection{Baseline VQE Energy}
\label{sec:vqe_energy}

Throughout this work the label \textbf{“VQE energy”} refers to the plain,
shot-based expectation value of the electronic Hamiltonian on the
approximate ground state $|\psi_\text{in}\rangle$  obtained in
Sec.~\ref{sec:ground_state_prep}.  Concretely, we evaluate
\begin{equation}
  E_{\mathrm{VQE}}
  \;=\;
  \sum_{i=1}^{K} w_i\,
  \bigl\langle \psi_\text{in} \bigl| P_i \bigr|
          \psi_\text{in} \bigr\rangle,
  \label{eq:vqe_energy}
\end{equation}
where \(H = \sum_i w_i P_i\) contains \(K=551\) Pauli strings.  The evaluation protocol depends on the backend:

\begin{itemize}
  \item \textit{Ideal simulator:}  
        Because an exact state vector is available, the inner products in
        Eq.~\eqref{eq:vqe_energy} are computed analytically, resulting in
        the single value reported in
        Table~\ref{tab:h2o_shot_energy_subspace} (no shot dependence).
  \item \textit{Noise-model simulation:}  
        The \(K\) strings are greedily packed into commuting groups
        (\(G=31\) groups total) and each group is sampled with
        \(N_g = 1\,000\) shots, yielding the aggregate
        \(\sum_g N_g = 31\,000\)-shot entry in
        Table~\ref{tab:h20_noisy_sim}.
  \item \textit{IBM Brisbane QPU:}  
        The same grouping is used, but each group receives
        \(N_g = 2\,000\) shots to reduce the variance in the energy, for a total of \(62\,000\) shots. Additionally, Twirled Readout Error Extinction (T-REX)~\cite{van2022model} was used to mitigate readout error, raising the total number of shots to \(\approx 200\,000\) performed to obtain the energy in Table~\ref{tab:h2o_shot_energy_subspace_real}.
\end{itemize}
All simulations and executions were carried out with \texttt{Qiskit 2.0}.
\subsection{QSCI}
\label{subsec:comp_qsci}

For the numerical experiments we do not diagonalize directly in the set
of the $R$ most probable determinants sampled from the trial state as in raw QSCI~\cite{kanno2023quantum}
(cf.\ Sec.~\ref{sec:qsci}). Instead we use an optimized version of QSCI, first partitioning every valid
bit-string into its spin-up (spin-$\alpha$) and spin-down (spin-$\beta$) halves and collecting the \emph{unique}
$\alpha_i$ and $\beta_j$ configurations.
\begin{align*}
  \mathcal U_{\alpha} \;&=\;
  \bigl\{\,
      \lvert \alpha_i \rangle
      \;\bigl\vert\;
      \text{$\alpha_i$ appears in at least one sampled determinant
            and } N_\alpha(\alpha_i)=N_\alpha
  \bigr\}\\
  \mathcal U_{\beta} \;&=\;
  \bigl\{\,
      \lvert \beta_j \rangle
      \;\bigl\vert\;
      \text{$\beta_j$ appears in at least one sampled determinant
            and } N_\beta(\beta_j)=N_\beta
  \bigr\}.
\end{align*}
(Here $N_\alpha$ and $N_\beta$ are the target spin populations which in our case $N_\alpha=N_\beta=4$; erroneous
bit-strings with the wrong electron count in either sector are discarded).
We then build the QSCI subspace with the cross product
\[
  \mathcal{S}_{\text{QSCI}}
  \;=\;
  \bigl\{
    \lvert \alpha_i \beta_j \rangle
    \;\bigl\vert\;
    \lvert \alpha_i \rangle \!\in\! \mathcal U_{\alpha},
    \;
    \lvert \beta_j \rangle \!\in\! \mathcal U_{\beta}
  \bigr\},
\]
which is guaranteed to conserve particle number \emph{and} $S_z$ by
construction. The projected Hamiltonian
$H_{\text{QSCI}}=\hat P_{\mathcal S_{\text{QSCI}}}\hat H
\hat P_{\mathcal S_{\text{QSCI}}}$
is then diagonalised in the usual fashion; the dimension of
$\mathcal{S}_{\text{QSCI}}$ is
$|\mathcal U_{\alpha}|\,|\mathcal U_{\beta}|$. This implementation is built on the \texttt{fci.kernel{\_}fixed{\_}space} method in PySCF~\cite{pyscf}.

\bigskip
\subsection{SQD}
\label{subsec:comp_sqd}

In our SQD experiments we start from the same spin-resolved pools
$\mathcal U_{\alpha}$ and $\mathcal U_{\beta}$ introduced above but then
\emph{merge} them into a single set of unique spin strings,
\[
  \mathcal U_{\text{tot}}
  \;=\;
  \mathcal U_{\alpha}\,\cup\,\mathcal U_{\beta}.
\]
The working subspace is the full cross product of this unified pool with
itself,
\[
  \mathcal{S}_{\text{SQD}}
  \;=\;
  \bigl\{\,\lvert u_i\,u_j\rangle
          \;\bigl\vert\;
          u_i,\,u_j \in \mathcal U_{\text{tot}}
  \bigr\},
\]
whose dimension is $|\mathcal U_{\text{tot}}|^2$.  Whenever the pair
$(u_i,u_j)$ produces the same determinant as $(u_j,u_i)$ (the
closed-shell case) we keep a single copy.  This procedure follows the motivation in Ref.~\cite{robledo2024chemistry}: sampling can return lone
open-shell determinants, e.g.\ $\lvert1001\rangle$, which by themselves
cannot form eigenstates of total-spin zero.  By including \emph{every}
compatible partner from $\mathcal U_{\text{tot}}$ (here
$\lvert0110\rangle$) the batch automatically spans both the open-shell
singlet
$\bigl(\lvert1001\rangle \pm \lvert0110\rangle\bigr)\!/\sqrt2$
and triplet combinations, eliminating spin contamination.

The Hamiltonian is then projected onto $\mathcal S_{\text{SQD}}$ and
diagonalized exactly as in QSCI; we retain the smallest
eigenvalue as the SQD energy estimate (see
Sec.~\ref{subsec:SQD}).
We compute the average occupancy of each orbital which we use to perform the \emph{self consistent configuration recovery}~\cite{robledo2024chemistry} subroutine to flip bits from the wrong (erroneous) configurations correcting them into configurations with the correct number of particles in each spin sector. We check the corrected configurations for new (not sampled previously) determinants which we include in the initial set and repeat the whole process for a given number of iterations growing the subspace size each time. The implementation of this algorithm was built on the \texttt{qiskit-addon-sqd} python package~\cite{qiskit-addon-sqd}.
\paragraph{Example:}
In the simplified case of a closed-shell molecule with two orbitals and two electrons, if the configurations $|1001\rangle$ and $|1010\rangle$ are sampled, we can separate $\mathcal U_{\alpha}=\{|10\rangle\}$ and $\mathcal U_{\beta}=\{|10\rangle, |01\rangle\}$. Feeding these sets to QSCI would yield $\mathcal{S}_{\text{QSCI}}= \{|1010\rangle, |1001\rangle\}$, while if we pass the same sets to SQD we will have $\mathcal{S}_{\text{SQD}}= \{|1010\rangle, |1001\rangle,  |0110\rangle, |0101\rangle\}$.

\section{Results}

\subsection{Ideal State-Vector Simulator}
\label{sec:ideal_results}

On the noise-free state-vector backend every gate is unitary and no
read-out or decoherence channels are present.  For the reference VQE we
therefore evaluate Eq.~\eqref{eq:vqe_energy} \emph{exactly} from the
state vector, the result is then a single,
shot independent value (\(-75.01247\,\text{Ha}\) in
Table~\ref{tab:h2o_shot_energy_subspace}).  In contrast, QSCI and SQD
are tested under finite measurement budgets: once the full probability
distribution has been generated from the converged state $|\psi_\text{in}\rangle$, we sample (draw)
\(N_{\text{shots}}\) samples to emulate measurement statistics.  Because
the Hartree–Fock determinant carries a weight of \(\approx 0.972\), 
both estimators stall at the lowest budget of $100$ samples because too few
distinct determinants are discovered. With \(10^{4}\) total samples the subspace
methods reach chemical accuracy, and at \(10^{6}\) samples—where the
determinant pool saturates at 225 configurations, the full configuration space, they reproduce the
exact energy.  Besides the optimization of the ground state parameters, all ideal-simulator runs combined required less than ten minutes of CPU time, so runtime was not a limiting factor in this experiment.

\begin{table}[H]
  \centering
  \setlength\tabcolsep{4pt}   
  \caption{Ground-state energies (Hartree) and subspace sizes for \ce{H2O} in the ideal case
            computed with different estimators under different shot budgets.  Energies that are within chemical accuracy ($\lvert\Delta E\rvert < 1.6\,\text{mHa}$) of the exact energy of the ground state are written in bold. VQE in the ideal case was estimated mathematically from the state vector representing the state and is not subject to varying shot budget.
            Subspace sizes that are marked with \protect\starlbl\  are equal to the full space and thus the computation is a Full Configuration Interaction computation.}
  \label{tab:h2o_shot_energy_subspace}

\begin{tabular}{@{}l *{5}{S[table-format=-2.5]}@{}}
  \cmidrule(lr){2-6}
  & \multicolumn{5}{c}{number of measurements} \\ \cmidrule(lr){2-6}
  & {$10^{2}$} & {$10^{3}$} & {$10^{4}$} & {$10^{5}$} & {$10^{6}$} \\[2pt]
  \midrule
  \textbf{VQE}                       & \multicolumn{5}{c}{\textbf{-75.01247}} \\
  \textit{Error} \(\lvert\Delta E\rvert\) & \multicolumn{5}{c}{0.00012} \\  
  \midrule
  \makecell{\textbf{number of}\\ \textbf{configurations discovered}}
                                     & {3} & {11} & {25} & {30} & {43} \\
  \midrule
  \multicolumn{6}{@{}l}{\textbf{QSCI}}\\[-4pt]
  \addlinespace[3pt]
  Subspace size                      & {9} & {42} & {100} & {121} & {225\starlbl} \\
  Energy                             & {-74.96742} & {-75.00049} & \textbf{-75.01233} & \textbf{-75.01248} & \textbf{-75.01259} \\
  \textit{Error} \(\lvert\Delta E\rvert\) & {0.04517} & {0.01210} & {0.00026} & {0.00011} & {0.00000} \\
  \midrule
  \multicolumn{6}{@{}l}{\textbf{SQD}}\\[-4pt]
  \addlinespace[3pt]
  Subspace size                      & {16} & {49} & {100} & {121} & {225\starlbl} \\
  Energy                             & {-74.98652} & {-75.00827} & \textbf{-75.01233} & \textbf{-75.01248} & \textbf{-75.01259} \\
  \textit{Error} \(\lvert\Delta E\rvert\) & {0.02607} & {0.00432} & {0.00026} & {0.00011} & {0.00000} \\
  \midrule
  Exact Energy                       & \multicolumn{5}{c}{-75.01259} \\
  \bottomrule
\end{tabular}
\end{table}


\subsection{Noise-Model Simulator}
\label{sec:noise_results}

\begin{figure}
    \centering
    \includegraphics[width=0.95\linewidth]{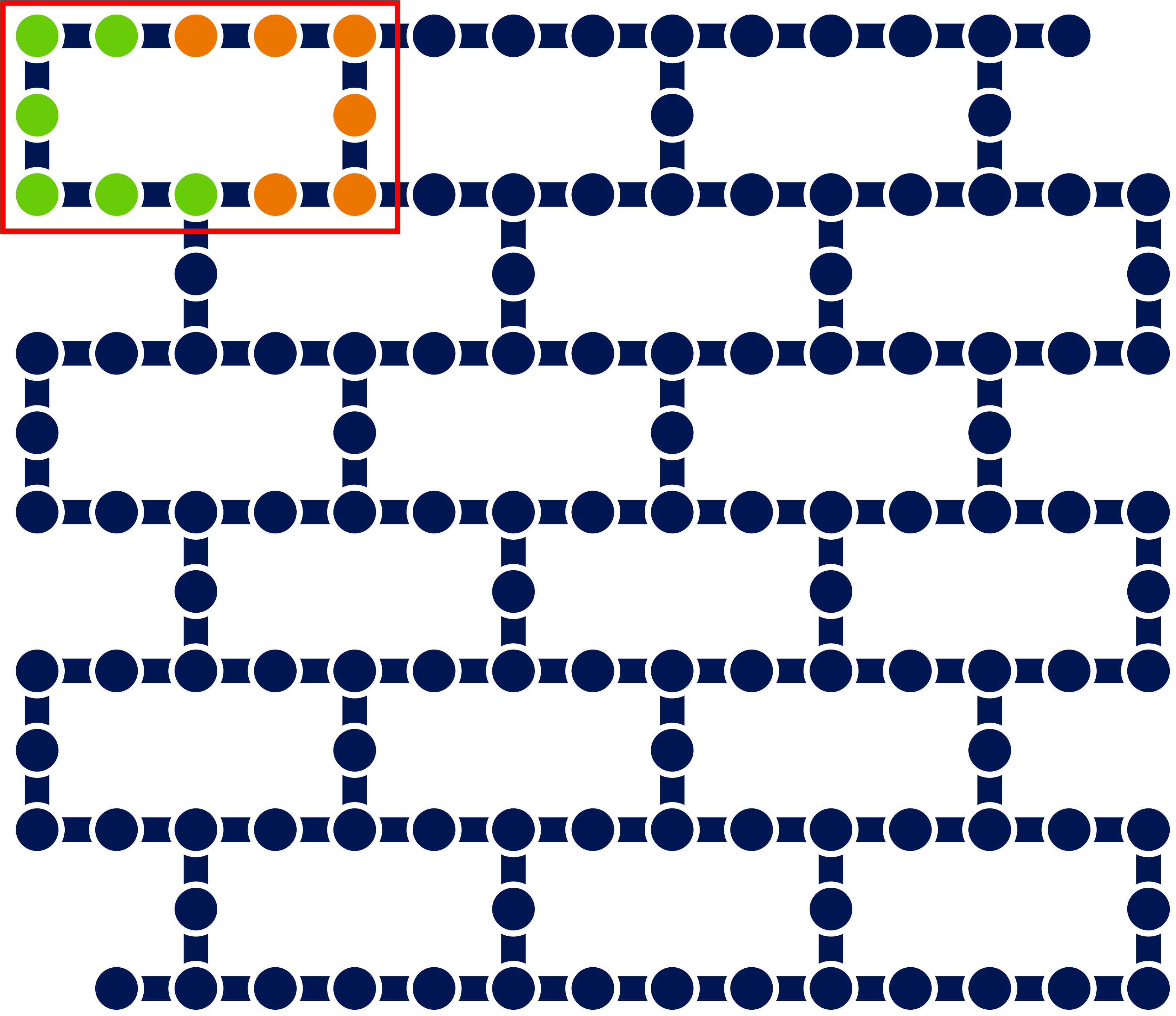}
    \caption{The \textsc{IBM Brisbane} QPU's qubit layout (vertices) and connectivity (edges). This QPU is of the Eagle family, comprised of $127$ qubits arranged in a heavy-hex lattice with cells of $12$ qubits. The entangling gates are $\operatorname{ECR}$ gates. The qubits used ( highlighted in the red-outlined cell), were manually selected based on their readout and $\operatorname{ECR}$ errors at the time of the execution. Green qubits designate the spin-$\alpha$ spin orbitals, and Orange qubits designate the spin-$\beta$ spin orbitals. The picture is adapted from IBM Quantum website~\cite{IBMQ2024} in accordance with applying terms.}
    \label{fig:brisbane_layout}
\end{figure}

To gauge robustness against realistic hardware errors we replaced the
state-vector backend with a full noise model of the \textsc{IBM Brisbane}
device calibrated on 1 June 2025.
The qubit-to-qubit connectivity, gate set and mapping exactly
match the layout later used on the real quantum processor
(see Fig.~\ref{fig:brisbane_layout}). The model includes amplitude- and phase-damping, two-qubit depolarizing noise, and read-out error.
After grouping the terms from the Hamiltonian into \(G=31\) commuting groups, each group was
measured with \(N_g = 1000\) shots (\(\sum_{g}^G N_g = 31\,000\)).  Under this noise, the VQE energy
drops to \(E_{\text{VQE}} = -69.98(23)\,\text{Ha}\),
with an error of \(\lvert\Delta E\rvert \approx 5.0\,\text{Ha}\)
(\(\sim 5\,000\,\text{mHa}\)), three orders of magnitude beyond chemical
accuracy.  QSCI and SQD, which sample determinants from the noisy state,
behave very differently.  With only \(10^{2}\) total samples SQD already
recovers the full 225-determinant space and delivers the exact energy,
while QSCI requires \(10^{4}\) shots to do the same, passing through an
intermediate 182-determinant stage where its error peaks at
\(0.404\,\text{Ha}\) (\(404\,\text{mHa}\)). Noise inflates the raw
determinant pool 98 $\rightarrow$ 875 $\rightarrow$ 3 719 sampled strings as the budget grows
but also boosts the count of \emph{correct} configurations used by the
subspace methods (6 $\rightarrow$ 46 $\rightarrow$ 207).  Wall-clock time reflects the larger
shot load of VQE: \(\sim\)131 min for the energy evaluation versus
\(\sim\)10 min for all QSCI/SQD data points; no error mitigation was applied at this stage.

\begin{table}[H]
  \centering
  \setlength\tabcolsep{4pt}
    \caption{Ground-state energies (Hartree) and subspace sizes for \ce{H2O}
            computed with different estimators under different shot budgets on a simulated noise model of the real quantum computer \textsc{IBM Brisbane}. Energies that are within chemical accuracy of the exact energy of the ground state are written in bold. The VQE energy was estimated with 1000 shots per group of observables resulting in a total of 31000 shots and is not subject to varying shot budget.
            Subspace sizes that are marked with \protect\starlbl\ are equal to the full space and thus the computation is a Full Configuration Interaction computation.}
    \label{tab:h20_noisy_sim}
  \begin{tabular}{@{}l *{3}{S[table-format=-2.5]}@{}}
    \cmidrule(lr){2-4}
        & \multicolumn{3}{c}{number of measurements} \\ \cmidrule(lr){2-4}
        & {$10^{2}$} & {$10^{3}$} & {$10^{4}$} \\[2pt]
    \midrule
    \textbf{VQE} & \multicolumn{3}{c}{-69.98454 $\pm$ 0.22742} \\
    \textit{Error} \(\lvert\Delta E\rvert\)
                 & \multicolumn{3}{c}{5.02805} \\      
    \midrule
    \textbf{number of configurations discovered} & {98} & {875} & {3719} \\
    \textbf{number of correct configurations}    & {6}  & {46}  & {207}  \\
    \midrule
    \multicolumn{4}{@{}l}{\textbf{QSCI}}\\[-4pt]
    \addlinespace[3pt]
    Subspace size  & {30} & {182} & {225\starlbl} \\
    Energy         & {-74.96274} & {-74.60852} & {\textbf{-75.01259}} \\
    \textit{Error} \(\lvert\Delta E\rvert\) 
                   & {0.04985} & {0.40407} & {0.00000} \\
    \midrule
    \multicolumn{4}{@{}l}{\textbf{SQD}}\\[-4pt]
    \addlinespace[3pt]
    Subspace size  & {225\starlbl} & {225\starlbl} & {225\starlbl} \\
    Energy         & {\textbf{-75.01259}} & {\textbf{-75.01259}} & {\textbf{-75.01259}} \\
    \textit{Error} \(\lvert\Delta E\rvert\)
                   & {0.00000} & {0.00000} & {0.00000} \\
    \midrule
    Exact Energy & \multicolumn{3}{c}{-75.01259} \\
    \bottomrule
  \end{tabular}
\end{table}

\subsection{Real Quantum Computer: \textsc{IBM Brisbane}}
\label{sec:hardware_results}

The final benchmark was executed directly on the 127-qubit
\textsc{IBM Brisbane} processor on 2 June 2025, using the \textbf{same
logical-to-physical qubit mapping and heavy-hex connectivity shown in
Fig.~\ref{fig:brisbane_layout}}.  Each of the \(G = 31\) commuting
measurement groups was sampled with \(N_g = 2000\) shots, and Twirled
Readout Error eXtinction (T-REX) was applied, bringing the effective
total to \(\sim 200\,000\) shots.  
Under these conditions the VQE returns
\(E_{\text{VQE}} = -69.89(7)\,\text{Ha}\),
an error of \(\lvert\Delta E\rvert \approx 5.12\,\text{Ha}\)
(\(\sim 5\,120\,\text{mHa}\)), in line with the noise-model prediction.
QSCI improves steadily with shot budget, reaching the exact energy once
its subspace saturates at 225 determinants, whereas SQD delivers the
exact result at every budget tested, thanks to its self consistent configuration recovery and multiple iterations.  
Job runtimes mirror the shot counts: the three QSCI/SQD jobs took
1 second, 4 seconds, and 12 seconds, respectively, while the VQE energy evaluation took
611 seconds of QPU time independently of queueing conditions.  
 The total quantum time consumed was \(\approx\)11 min.
No additional error-mitigation or post-selection was applied beyond the
T-REX readout correction.

\begin{table}[H]
  \centering
  \setlength\tabcolsep{4pt}
 \caption{Ground-state energies (Hartree) and subspace sizes for \ce{H2O}
            computed with different estimators under different shot budgets on the real quantum computer \textsc{IBM Brisbane}. Energies that are within chemical accuracy ($\lvert\Delta E\rvert < 1.6\,\text{mHa}$) of the exact energy of the ground state are written in bold. The VQE energy was estimated with 2000 shots per group of observables resulting in a total of 62000 shots and is not subject to varying shot budget.
            Subspace sizes that are marked with \protect\starlbl\ are equal to the full space and thus the computation is a Full Configuration Interaction computation.}
  \label{tab:h2o_shot_energy_subspace_real}

  \begin{tabular}{@{}l *{3}{S[table-format=-2.5]}@{}}
    \cmidrule(lr){2-4}
        & \multicolumn{3}{c}{number of measurements} \\ \cmidrule(lr){2-4}
        & {$10^{2}$} & {$10^{3}$} & {$4\times10^{3}$} \\[2pt]
    \midrule
    \textbf{VQE} & \multicolumn{3}{c}{-69.89190 $\pm$ 0.07174} \\
    \textit{Error} $\lvert\Delta E\rvert$
                 & \multicolumn{3}{c}{5.12069} \\   
    \midrule
    \textbf{number of configurations discovered} & {99} & {857} & {2330} \\
    \textbf{number of correct configurations}    & {6}  & {41}  & {109}  \\
    \midrule
    \multicolumn{4}{@{}l}{\textbf{QSCI}}\\[-4pt]
    \addlinespace[3pt]
    Subspace size  & {25} & {210} & {225\starlbl} \\
    Energy         & {-73.82242} & {-75.00679} & {\textbf{-75.01259}} \\
    \textit{Error} $\lvert\Delta E\rvert$
                   & {1.19017} & {0.00580} & {0.00000} \\
    \midrule
    \multicolumn{4}{@{}l}{\textbf{SQD}}\\[-4pt]
    \addlinespace[3pt]
    Subspace size  & {225\starlbl} & {225\starlbl} & {225\starlbl} \\
    Energy         & {\textbf{-75.01259}} & {\textbf{-75.01259}} & {\textbf{-75.01259}} \\
    \textit{Error} $\lvert\Delta E\rvert$
                   & {0.00000} & {0.00000} & {0.00000} \\
    \midrule
    Exact Energy & \multicolumn{3}{c}{-75.01259} \\
    \bottomrule
  \end{tabular}
\end{table}

\section{Discussion of Key Findings}
\label{sec:discussion}

In the noiseless simulator the Hartree–Fock determinant dominates the
probability distribution of the approximate ground state with a weight
of \(0.972\).
As a result, successive measurements overwhelmingly resample this single
configuration, making the discovery of new determinants extremely slow:
at \(10^{2}\) total shots only 3 unique strings are found, and even at
\(10^{6}\) shots the total rises to just 43
(Tab.~\ref{tab:h2o_shot_energy_subspace}).
This behavior mirrors the sampling bottleneck analyzed in
Chapter~\ref{chap3} and agrees with the predictions of Reinholdt
\emph{et al.}\,\cite{reinholdt2025exposing}. A \(10^{4}\) fold increase in the number of samples (\(10^{2} \rightarrow 10^{6}\)) only increased the number of discovered determinants by a factor of $\approx14$ (\(3 \rightarrow 43\)). This highlights the fact that
determinant discovery is the closer asymptotic bottleneck of any
sampling based algorithms instead of diagonalization.

When the shot budget crosses \(10^{4}\), both QSCI and SQD generate
$\ge 100$ distinct configurations, their subspaces then
estimate the ground state energy within \(0.26\ \mathrm{mHa}\).
With \(10^{6}\) samples, the subspace saturates at 225 determinants, the full
configuration space, and the estimated energy matches the FCI benchmark.

Introducing the calibrated \textsc{IBM Brisbane} noise model inflates the raw
measurement pool from 98 to 3719 bit-strings across the three shot
budgets, and, crucially, boosts the count of \emph{valid} determinants
from 6 to 207
(Tab.~\ref{tab:h20_noisy_sim}).
SQD capitalizes on this immediately: with only \(10^{2}\) samples its
inflated subspace already spans all 225 determinants and returns
the exact energy.
QSCI benefits more gradually, converging at \(10^{4}\) shots after
passing through an intermediate 182-determinant stage where its error
briefly reaches \(404\ \mathrm{mHa}\).
These observations suggest that stochastic device noise can act as a
\emph{diversification mechanism}, seeding the subspace with otherwise
rare yet chemically important configurations.

Running the same protocols on the real 127-qubit \textsc{IBM Brisbane} processor
reproduces the simulator hierarchy: SQD is chemically accurate at every
budget; QSCI converges once its subspace is full; and the measurement
cost scales exactly with the overhead predicted in Chapter~\ref{chap3}.
The three QSCI/SQD jobs consume a total of 17 s of QPU time, whereas the
VQE energy evaluation, dominated by the readout error correction overhead in addition to the 62 000 commuting group measurements, requires 611 s.
Expressed in the common metric of “mHa per quantum-second,”
quantum subspace diagonalization is therefore at least two orders of
magnitude more resource efficient than conventional VQE on current
hardware.

\smallskip
\noindent
Collectively, these findings demonstrate the potential of quantum sampling methods combined with symmetry optimizations and error mitigation techniques. Their ability to estimate chemically accurate energies with minimal shot budgets compared to VQE validates the methodological propositions advanced by Robledo-Moreno \emph{et al.}~\cite{robledo2024chemistry}.

\chapter{Conclusion}

Accurately predicting electronic ground‐state energies remains a flagship target for quantum advantage in computational chemistry.  
This thesis therefore evaluates and explores \emph{sampling‐based} quantum–classical algorithms—namely, \textbf{Quantum‐Selected Configuration Interaction} (QSCI) and \textbf{Sample‐Based Quantum Diagonalization} (SQD)—as near-term successors to the Variational Quantum Eigensolver (VQE).  
The study is framed within the emerging paradigm of \emph{quantum-centric supercomputing}, in which quantum processors tackle those parts of a workflow that benefit most from quantum principles, while high-performance classical resources complete the remaining tasks.

A novel connection has been established between the determinant discovery step in quantum sampling-based algorithms and the classical \emph{coupon-collector} problem.  
This analysis yields (i) an exact expression for the expected number of measurements required to uncover all relevant determinants and (ii) a tractable lower-bound approximation suitable for large determinant sets. This connection shows that the problem facing sampling is fundamental and not a product of NISQ hardware characteristics. 
In addition, state-of-the-art protocols for QSCI and SQD have been embedded in a unified workflow using \texttt{Qiskit}~2.0~\cite{qiskit}, incorporating symmetry optimizations and error mitigation, all fully compatible with present-day quantum hardware.

All simulations and hardware experiments were performed by the author.  
The algorithms were tested on the water molecule (\ce{H2O}, STO-3G basis set, 12-qubit active space) under three progressively realistic scenarios:  
(i) ideal state-vector simulation,  
(ii) hardware-calibrated noisy simulation, and  
(iii) execution on IBM’s 127-qubit quantum computer \textsc{IBM Brisbane}.  
Across all regimes SQD reached chemical accuracy while requiring orders of magnitude fewer shots than VQE; its classical post-processing amplifies the information gleaned per quantum measurement.   
Empirical shot counts closely track the coupon-collector predictions, validating the theoretical framework and confirming that discovery of rare determinants is the dominant cost driver.

Despite the encouraging results, several challenges remain in the way of quantum advantage.  
First, as discussed in Chapter~\ref{chap3} and in the numerical study of Reinholdt \emph{et~al.}~\cite{reinholdt2025exposing}, the \emph{sampling phase} is still the principal bottleneck: discovering determinants with very low amplitudes demands a rapidly growing number of measurements as the probability distribution becomes increasingly skewed, i.e in the weakly correlated regime, exactly as predicted by the coupon–collector analysis.  
Second, once those measurements have been gathered, the ensuing classical diagonalization scales quadratically with the size of the selected subspace ($\mathcal{O}(d^{2})$ in both memory and time); if the determinant pool expands unchecked, this cost can quickly dominate the overall workflow. Although this problem is not exclusive to quantum methods, classical state of the art methods like the Heat-Bath Configuration Interaction algorithm~\cite{holmes2016heat} handle it better by considering important determinants first.
Finally, the present implementation relies on relatively straightforward error-mitigation strategies that leave ample room for improvement; more sophisticated techniques, together with smarter sampling, will be necessary to sustain these algorithms as molecular systems grow, and eventually outperform classical methods.

Faster, lower-noise hardware and smarter sampling strategies offer a clear path toward closing the performance gap with classical benchmarks and realizing quantum advantage, especially in strongly correlated problems.  
The coupon-collector framework introduced here provides a quantitative yardstick for such improvements, while the experimentally validated QSCI/SQD recipes demonstrate that advanced quantum chemistry protocols can already be deployed on NISQ devices. This marks an important milestone for quantum-centric supercomputing and quantum computing in general.

\section*{Future Directions}
Looking ahead, several research avenues emerge naturally:
\begin{itemize}
  \item \textbf{Amplitude-flattening ansätze} or \textbf{state-preparation heuristics} that reduce sampling skew and accelerate rare-determinant discovery.
  \item \textbf{Importance-sampling and active-learning} techniques to steer measurements toward unexplored regions of determinant space.
  \item Hybrid \textbf{SCI $\leftrightarrow$ SQD} strategies that exploit the complementary strengths of classical state-of-the-art algorithms and quantum sampling-based ones.
\end{itemize}
With these directions in view, this dissertation contributes a rigorous theoretical toolset, a validated experimental workflow, and a realistic assessment of present capabilities, laying the groundwork for the next generation of quantum algorithms in electronic-structure theory.
\pagebreak
\bibliography{bib} 

\end{document}